\documentclass[a4paper,11pt]{article}
\pdfoutput=1 %

\usepackage{jheppub} 
\usepackage{amsmath,amssymb,graphicx,float,slashed,xcolor,multicol}
\usepackage{tabularx}
\usepackage{url}
\usepackage{footmisc}
\usepackage{amsfonts}
\usepackage{cancel}
\usepackage{color}
\usepackage{multirow} 
\usepackage{pifont}
\usepackage{epstopdf}
\usepackage{comment}
\usepackage{booktabs} 
\usepackage{natbib}
\usepackage{array}
\usepackage{mathrsfs}
\usepackage[toc,page]{appendix}
\usepackage{mathtools}
\usepackage{romannum}
\usepackage{ulem}
\usepackage{bbold}
\usepackage{enumitem}
\usepackage{multirow}
\usepackage{cleveref}
\usepackage{caption,subcaption}
\usepackage{float}
\usepackage{multirow}
\usepackage{upgreek}
\usepackage[toc,page]{appendix}
\usepackage{romannum}
\allowdisplaybreaks
\usepackage{bbm}                
\usepackage{xspace}				

\usepackage{tikz}
\usetikzlibrary{arrows,shapes}
\usetikzlibrary{trees}
\usetikzlibrary{matrix} 
\usetikzlibrary{positioning}				
\usetikzlibrary{calc,through}				
\usetikzlibrary{decorations.pathreplacing}  
\usepackage{pgffor}							
\usetikzlibrary{decorations.pathmorphing}	
\usetikzlibrary{decorations.markings}

\tikzstyle{block} = [draw, rectangle, 
minimum height=3em, minimum width=6em]

\newcommand*{\rom}[1]{\expandafter\@slowromancap\romannumeral #1@}

\newcommand{\cO}{\mathcal{O}}
\newcommand{\cL}{\mathcal{L}}
\newcommand{\cM}{\mathcal{M}}
\newcommand{\cP}{\mathcal{P}}

\newcommand{\fb}{\mathrm{fb}}

\newcommand{\BR}{\mathrm{BR}}

\newcommand{\mgg}{m_{\gamma\gamma}}
\newcommand{\Dmgg}{\Delta \mgg}

\newcommand{\GeV}{\mathrm{GeV}}
\newcommand{\TeV}{\mathrm{TeV}}

\newcommand{\fm}{\mathrm{fm}}

\newcommand{\eg}{\textit{e.g.}}
\newcommand{\ie}{\textit{i.e.}}

\newcommand{\abs}[1]{| #1 |}

\def\beq{\begin{equation}}
\def\eeq{\end{equation}}
\def\beqa{\begin{eqnarray}}
\def\eeqa{\end{eqnarray}}

\title{Probing axion-like particles at the Electron-Ion Collider}
\author[a,g]{Reuven Balkin,}
\author[b]{Or Hen,}
\author[c,d]{Wenliang Li,}
\author[a]{Hongkai Liu,}
\author[a,e,f]{Teng Ma,}
\author[a]{Yotam Soreq,}
\author[b]{Mike Williams}

\affiliation[a]{Physics Department, Technion – Israel Institute of Technology, Haifa 3200003, Israel}
\affiliation[b]{Laboratory for Nuclear Science, Massachusetts Institute of Technology, Cambridge, MA 02139, USA}
\affiliation[c]{Center for Frontiers in Nuclear Science, Stony Brook, 11794, NY, USA}
\affiliation[d]{Stony Brook University, Stony Brook, 11794, NY, USA}
\affiliation[e]{International Centre for Theoretical Physics Asia-Pacific (ICTP-AP),
University of Chinese Academy of Sciences (UCAS), 100190 Beijing, China}
\affiliation[f]{IFAE and BIST, Universitat Aut\`onoma de Barcelona, 08193 Bellaterra, Barcelona}
\affiliation[g]{Department of Physics, University of California Santa Cruz and Santa Cruz Institute for Particle Physics, 1156 High St., Santa Cruz, CA 95064, USA}

\emailAdd{
rebalkin@ucsc.edu, 
hen@mit.edu,
billlee@jlab.org,
liu.hongkai@campus.technion.ac.il, 
mateng@ucas.ac.cn,
soreqy@physics.technion.ac.il,
mwill@mit.edu}

\abstract{
The Electron-Ion Collider~(EIC), a forthcoming powerful high-luminosity facility, represents an exciting opportunity to explore new physics.
In this article, we study the potential of the EIC to probe the coupling between axion-like particles~(ALPs) and photons in coherent scattering.
The ALPs can be produced via photon fusion and decay back to two photons inside the EIC detector. 
In a prompt-decay search, we find that the EIC can set the most stringent bound for $m_a \lesssim 20\,\GeV$ and probe the effective scales $\Lambda \lesssim 10^{5}\,$GeV.
In a displaced-vertex search, which requires adopting an EM calorimeter technology that provides directionality, the EIC could probe ALPs with $m_a \lesssim 1\,\GeV$ at effective scales $\Lambda
\lesssim 10^{7}\,\GeV$. 
Combining the two search strategies, the EIC can probe a significant portion of unexplored parameter space in the $0.2 < m_a <20\,\GeV$ mass range.
} 

\begin{document}
	
\titlepage
\maketitle

\flushbottom

\section{Introduction}
\label{sec:intor}

The Electron-Ion Collider~(EIC)~\cite{Accardi:2012qut,AbdulKhalek:2021gbh} at Brookhaven National Laboratory, which will be the first high-intensity lepton-ion collider of its kind, is expected to start taking data during the next decade. 
It will collide high-energy electrons with protons or ions, with maximal electron and ion energies of up to 18\,GeV and 275\,GeV per nucleon, respectively. 
Depending on the running mode, the integrated luminosity could reach $\sim100\,\fb^{-1}$ for electron-ion collisions. 
In addition to its rich Standard Model~(SM) program, which focuses on nuclear and hadronic physics, the EIC has the potential to explore new physics beyond the SM~(BSM), \eg~\cite{Gonderinger:2010yn, Cirigliano:2021img,Davoudiasl:2021mjy,Zhang:2022zuz,Batell:2022ogj,Yan:2022npz,AbdulKhalek:2022hcn,Davoudiasl:2023pkq,Boughezal:2022pmb,Liu:2021lan,Yue:2023mew,Wang:2024zns}.

One well-motivated BSM scenario is the QCD axion and more generally axion-like particles~(ALPs). 
The QCD axion was originally introduced as part of the Peccei-Quinn mechanism, a proposed solution to the strong CP problem~\cite{Peccei:1977hh,Peccei:1977ur,Weinberg:1977ma,Wilczek:1977pj}, while ALPs appear generically in many 
frameworks \eg\ string theory~\cite{Witten:1984dg,Svrcek:2006yi,Conlon:2006tq}. 
These ALPs can serve as portals to dark sectors~\cite{Nomura:2008ru,Freytsis:2010ne,Dolan:2014ska,Hochberg:2018rjs,Ghosh:2023tyz,Dror:2023fyd,Fitzpatrick:2023xks}, and if light enough, as viable dark matter candidates themselves~\cite{Preskill:1982cy,Abbott:1982af,Dine:1982ah}.

In this work, we study the potential of the EIC to probe ALPs with a photon coupling in the sub-GeV to $\cO(20)\GeV$ mass range. 
We focus on the coherent production of ALPs (\ie{} in which the ion stays intact) due to several reasons. 
First, the coherent production based on electromagnetic processes has a $Z^2$-enhanced cross section, where $Z$ is the atomic number of the ion.  
Second, as opposed to fixed target experiments in which the coherent cross section decouples above the GeV scale, see \eg~\cite{Aloni:2019ruo}, the boosted ion allows coherent production of ALPs with masses up to $\cO(20)\,\GeV$. 
Lastly, the requirement for coherent events significantly reduces the amount of background events.

In the region of parameter space where the ALP decays promptly to two photons close to the interaction point,
we find that the EIC has the potential to uniquely probe unexplored ALP parameter space below masses of 20\,GeV and effective scales of $\Lambda \sim 10^{5}\,\GeV$.
In addition, we consider the possibility of displaced ALP decays inside the detector. 
Assuming the EIC detector has sufficient diphoton-vertex resolution, we find that it can potentially probe additional unexplored regions of parameter space in the sub-GeV mass range, with sensitivity which is comparable to other proposals.

There are numerous experimental bounds on sub-GeV ALPs from various terrestrial searches, such as
beam dumps and kaon experiments~\cite{CHARM:1985anb,Riordan:1987aw, Bjorken:1988as,Blumlein:1990ay,Dobrich:2015jyk,Dobrich:2017gcm,Harland-Lang:2019zur,Dobrich:2019dxc,NA64:2020qwq,Afik:2023mhj,Ema:2023tjg},
$B$-factories~\cite{Belle-II:2020jti}, photon beams~\cite{Feng:2018pew,Aloni:2019ruo,Balkin:2021jdr,GlueX:2021myx,Pybus:2023yex}, LEP~\cite{OPAL:2002vhf,Jaeckel:2015jla,Yue:2021iiu,Tian:2022rsi,Bao:2022onq,BESIII:2022rzz} 
and the LHC~\cite{CMS:2012cve,ATLAS:2014jdv,Mimasu:2014nea,ATLAS:2015rsn,Brivio:2017ije, Bauer:2017ris, Ebadi:2019gij,Bonilla:2022pxu,Alonso-Alvarez:2023wni,Mitridate:2023tbj,Dutta:2023abe}.
A related type of search for ALPs using ultra-relativistic ions, based on ultra-peripheral heavy-ion collisions at the LHC~\cite{Knapen:2016moh,Knapen:2017ebd}, was recently performed by CMS~\cite{CMS:2018erd} and ATLAS~\cite{ATLAS:2020hii}. 
While the coherent production in these searches is $Z^4$-enhanced and recovering both ions intact ensures low background rates, the integrated luminosity for heavy ion collision at the LHC is expected to be much smaller than the electron-ion luminosity at the EIC.
Finally, there are also several proposed experiments and searches aimed at probing the ALP parameter space in the future~\cite{SHiP:2015vad,Berlin:2018pwi,Bai:2021gbm,RebelloTeles:2023uig}. 

The rest of this paper is organized as follows. 
In Section~\ref{sec:EICdet}, we provide a brief description of the EIC detector. 
In Section~\ref{sec:coherentprod}, we discuss the production of ALPs at the EIC through coherent scattering. 
After providing the details of our prompt search and displaced-vertex search strategies, we present the corresponding projected sensitivities in the ALP parameter space.  
Finally, we present our conclusions and outlook in Section~\ref{sec:outlook}.
Technical details are provided in Appendices~\ref{sec:xs}.
In addition, the EIC sensitivity to dark photons is briefly discussed in Appendix~\ref{sec:dark_photons}.

\section{The Electron-Ion Collider}
\label{sec:EICdet}
 
The EIC detector~\cite{EPICurl} is being built by the ePIC collaboration based on a reference design developed by the ECCE consortium~\cite{Adkins:2022jfp} and the ATHENA collaboration~\cite{ATHENA:2022hxb}. 
It will be located in the IP6 location
at the Relativistic Heavy Ion Collider~(RHIC)~\cite{RHICurl} in Brookhaven National Laboratory, which currently hosts the STAR experiment~\cite{STARurl}. 
The experiment consists of a central detector, far-forward spectrometer, and far-backward spectrometer. 
The forward region is defined as the ion beam direction, while the backward region refers to the electron beam direction. 

The EIC central detector has a cylindrical geometry and is equipped with a solenoid which produces a 1.7\,T magnetic field, and is divided into three sectors covering different $\eta$ ranges: 
(i)~the barrel (pseudorapidity coverage $-1.7 < \eta < +1.3$), 
(ii)~the forward endcap ($+1.3 < \eta < +3.5$), and 
(iii)~the backward endcap ($-3.5 < \eta < -1.7$).
Each of the sectors is equipped with detectors for particle identification and calorimetry to tag final state particles, see~\cite{Adkins:2022jfp} for more details.  
In this work we focus on a diphoton final state, making the electromagnetic calorimeters the most critical components for our search. 
The acceptance and design parameters for the barrel, forward, and electron-end-cap electromagnetic calorimeters (BEMC, FEMC, and EEMC, respectively) are taken from~\cite{Bock:2022lwp} and are summarized in Table~\ref{tab:energyresolution}.
\begin{table}[t]   
\begin{center}    
\begin{tabular}{|c|c|c|}   
\hline   
\textbf{Calorimeter} & 
\textbf{Pseudorapidity acceptance} & 
\textbf{Projected energy resolution ($\Delta E/E)$ [\%]} \\   
\hline  
FEMC & 
$[+1.3\,,\,+3.5]$ & 
$7.1/\sqrt{E/\text{GeV}}$  \\  
\hline  
BEMC & 
$[-1.7\,,\,+1.3]$ & 
$1.6/\sqrt{E/\text{GeV}} \oplus 0.7$  \\ 
\hline   
EEMC & 
$[-3.5\,,\,-1.7]$ & 
$1.8/\sqrt{E/\text{GeV}} \oplus 0.8 $ \\
\hline   
\end{tabular} 
\caption{The acceptances and projected calorimeter resolutions used in the simulation study. These are based on information provided in Table~6 of Ref.~\cite{Adkins:2022jfp} and Ref.~\cite{Bock:2022lwp}. The resolutions used in this study represent the typical behavior of the calorimeters, obtained by averaging the best and worst scenarios reported in Ref.~\cite{Bock:2022lwp}. }
\label{tab:energyresolution} 
\end{center}   
\end{table}
The lab frame beam energies considered in this work are\footnote{We assume the beams are unpolarized since polarization does not play a significant role in the ALP production.}
\begin{align}
    E_e = 18\,\GeV   \quad \text{and} \quad
    E_{\rm Pb} = 20\, \TeV\,, 
\end{align}
or equivalently $\sim 100\,\GeV$ per nucleon. 
We consider two benchmark EIC integrated luminosities of $\mathcal{L} = 10\,\fb^{-1}$ and $ \mathcal{L} =100\,\fb^{-1}$.

\section{ALPs at the EIC}
\label{sec:coherentprod}

\subsection{Coherent production}
\label{sec:ALP_production}

\begin{figure}[t]
	\centering
\raisebox{-0.5\height}{\includegraphics[width=0.3\textwidth]{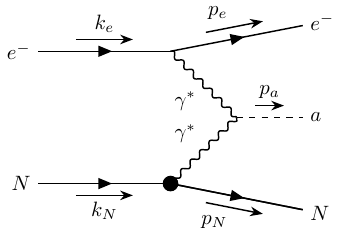}}
\raisebox{-0.5\height}{ \includegraphics[width=0.55
\textwidth]{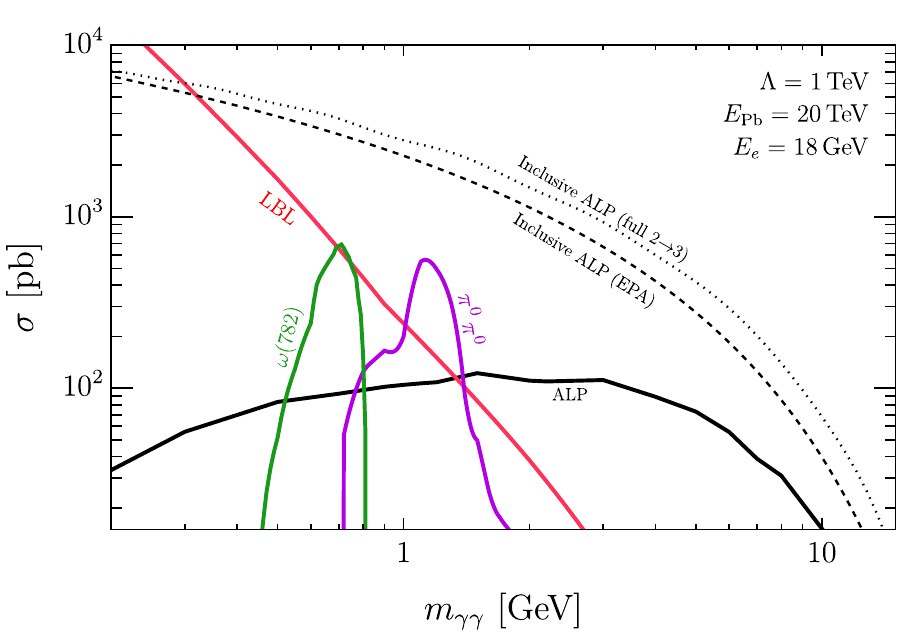}}
	\caption{
    Left panel: Feynman diagram of photon fusion production of an ALP at the EIC. 
    Right panel: 
    various cross sections as a function of the final state diphoton mass. The solid curves show the cross sections of both the signal ($\Lambda=1\,$TeV) and backgrounds after applying all the cuts discussed in Section~\ref{sec: prompt}. 
    The inclusive ALP production cross section is plotted in dashed\,(dotted) black for $\Lambda=1\,$TeV using the full $2~\to~3$\,(EPA) calculation.
    For the backgrounds $\sigma \approx 4\Dmgg d\sigma/d\mgg$ is plotted, where $\Dmgg$ is the invariant mass resolution, which is given in Table~\ref{tab:mgg_res}.  
    }	
\label{fig:feyn_and_xsecs}
\end{figure}

We start by discussing the coherent production of ALPs at the EIC.
We focus on massive ALPs, denoted by $a$, which interact predominantly with photons,\footnote{The investigation of interactions with other SM fields, such as gluons and fermions, is left for future work.}
\begin{align}
    \label{eq:lag}
    \cL_{a}
=   -\frac12 m_a^2 a^2- \frac{a}{4\Lambda} F^{\mu\nu}\tilde{F}_{\mu\nu},
\end{align}
where $\tilde{F}_{\mu\nu}\equiv({1}/{2})\epsilon_{\mu\nu\rho\sigma}F^{\rho\sigma}$.
We assume $\Lambda \gg m_a$ due to the pseudo-Goldstone nature of the ALP. 
The ALP is mainly produced via the 2-to-3 process,
\begin{align}
    \label{eq:eN2eNa}
    e^- (k_e) + N (k_N) \rightarrow e^- (p_e) + N (p_N)  + a (p_a) \, ,
\end{align}
where $N$ denotes the ion. 
The leading contribution to Eq.~\eqref{eq:eN2eNa} comes from photon fusion, see the left panel of Fig.~\ref{fig:feyn_and_xsecs}.

The differential cross section of $e^-N\to e^- N a$ naively depends on five independent kinematical variables, as well as on the (fixed) center-of-mass energy. 
However, due to the rotational symmetry of the initial state around the beam axis, one azimuthal angle dependence in the final state can be removed, making its integration trivial. 
Therefore, we define the following four Lorentz invariant variables:  
\begin{align}
    \label{eq::lorentz_inv_notation}
    s \equiv \left(k_e + k_N\right)^2 \, , \quad 
    t_e \equiv \left(k_e - p_e\right)^2 \, , \quad
    t_N \equiv \left(k_N - p_N\right)^2 \, , \quad
    m_{aN}^2 \equiv \left( p_a + p_N \right)^2 \, ,
\end{align}
and the angle 
\begin{align}
    \cos\theta \equiv 
    \frac{
    (\vec{k}_N \times \vec{p}_N ) \cdot
    ( \vec{k}_e \times \vec{p}_e )}
    {\abs{\vec{k}_N \times \vec{p}_N} \abs{\vec{k}_e \times \vec{p}_e}} \, ,
\end{align}
where $\theta$ is defined in the rest frame of the outgoing $a+N$ system. 
It can be understood as the angle between the planes spanned by $\{\vec{k}_N,\vec{p}_N\}$ and by $\{\vec{k}_e, \vec{p}_e\}$.
From the above definitions, we identify $t_e$ as the electron-transferred momentum and $t_N$ as the ion-transferred momentum. 
The differential cross section is 
\begin{align}
    \label{eq:dsig23}
    \frac{d \sigma^{2\to 3}_{a}}{d t_e dt_N dm^2_{aN} d\theta}
=   \frac{1}{(2\pi)^4}\frac{1}{4\sqrt{\lambda(s,m_e^2,m_N^2)}}
    \frac{1}{4\sqrt{\lambda(m_{aN}^2,m_N^2,t_e)}}
    \frac{\abs{\cM^{2\to 3}_a}^2}{4[(k_e\cdot k_N)^2-m_e^2m_N^2]^{1/2}} \, ,
\end{align}
where $\abs{\cM^{2\to 3}_a}^2 \propto (Z^2e^4)/(t_e^2t_N^2 \Lambda^2_a)$ is the $e^-N\to e^-N a$ matrix element squared and $\lambda(a,b,c)\equiv a^2 + b^2+c^2-2ab-2ac-2bc$. 
More details on the calculation of the matrix element are given in Appendix~\ref{sec:xs}. 
The phase space integration follows the method in Ref.~\cite{Kersevan:2004yh}.

Due to the double pole structure of the amplitude, the cross section is dominated by regions in parameter space where $|t_e|$ and $|t_N|$ are minimized.
The minimal momentum transfers are constrained by kinematics, and in the $m_e,\, m_a \ll m_N\ll \sqrt{s}$ limit they are given approximately by
\begin{align}
    \label{eq:tmin}
    |t_e|_{\rm min} &\approx 1.9\times 10^{-14}~\text{GeV}^2 \left(\frac{m_a}{1.0~\text{GeV}}\right)^2\left(\frac{m_N}{193~\text{GeV}}\right)^2
    \left(\frac{\sqrt{s}}{1.2~\text{TeV}}\right)^{-4}\,,\nonumber\\
    |t_N|_{\rm min} &\approx  1.8\times 10^{-8}~\text{GeV}^2 \left(\frac{m_a}{1.0~\text{GeV}}\right)^4\left(\frac{m_N}{193~\text{GeV}}\right)^2\left(\frac{\sqrt{s}}{1.2~\text{TeV}}\right)^{-4}\,.
\end{align}
The dominance at small momenta transfers implies that the calculation of the ALP photoproduction cross section could be greatly simplified by using the equivalent photon approximation~(EPA), see \eg~\cite{Dobrich:2015jyk}. 
However, as discussed in Sec.~\ref{sec:background}, the kinematics of the outgoing electron are required in order to veto background events. 
Thus, we calculate the full $2\to3$ process to retain the recoil information of the electrons. 
In the right panel of Fig.~\ref{fig:feyn_and_xsecs}, we present the total production cross section obtained from the full $2\to 3$ calculation~(dashed), which can be compared to the one obtained from the EPA~(dotted). 
They are in good agreement with each other especially for light ALP masses, while for heavier ALP masses they differ by about a factor of $2$ due to the breakdown of EPA at larger momentum transfers.
\begin{figure}[t] 
\includegraphics[width=0.48\textwidth]{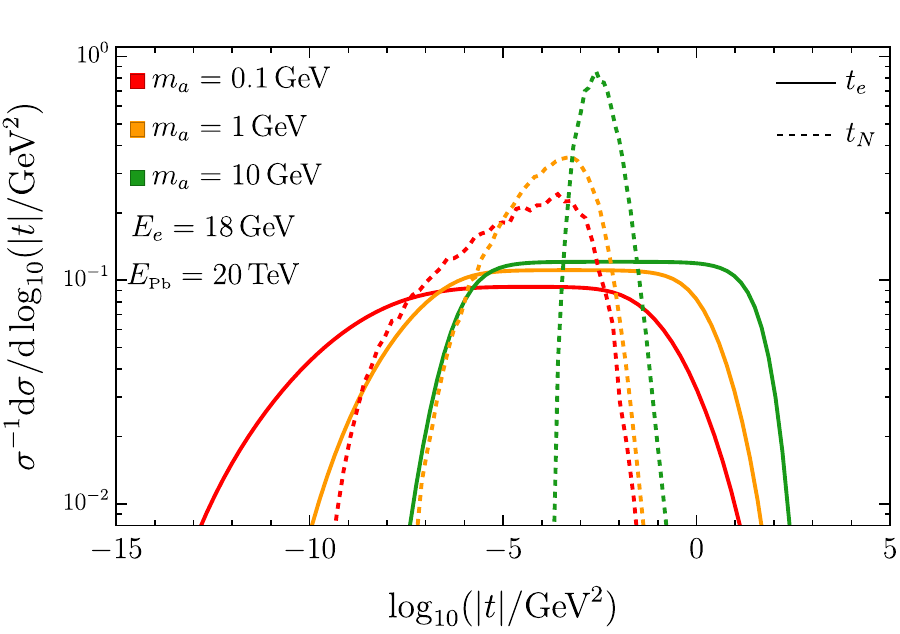}
\includegraphics[width=0.48
\textwidth]{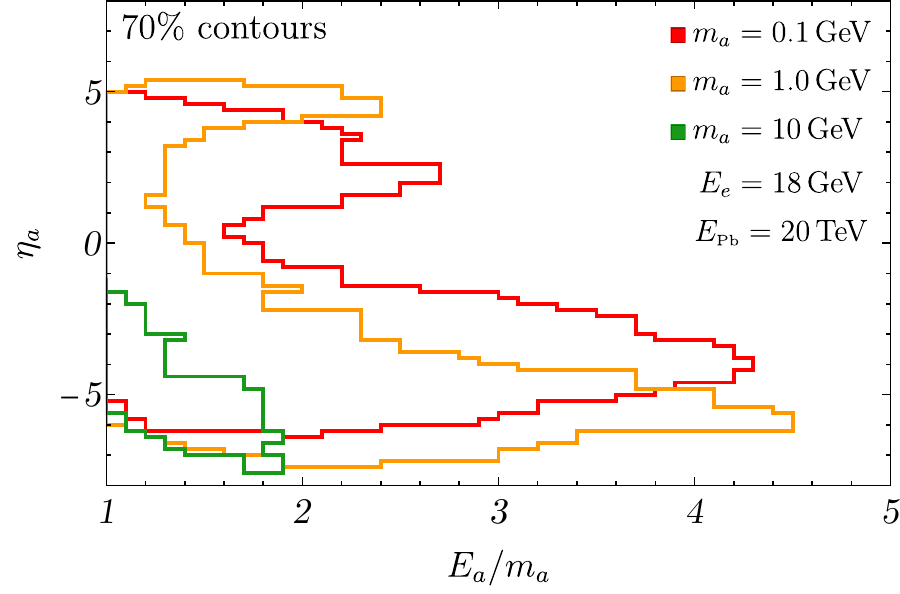}
	\caption{
	 Left panel: Normalized $\mathrm{d}\sigma/\mathrm{d}\log|t_e|$ (solid) and $\mathrm{d}\sigma/\mathrm{d}\log|t_N|$ (dashed) distributions of the ALP production cross section for $m_a =0.1, 1.0$ and $10.0\,$GeV plotted in red, orange and green, respectively. Right panel: The kinematical properties of the produced ALP represented by contours containing $70\,\%$ of the 2D probability distribution in the $\{E_a/m_a,\eta_a\}$ plane. 
  }	
	\label{fig:diff_and_boost}
\end{figure}

The $Z^2$-enhanced coherent production is suppressed when the transferred momentum to the nucleus is of the order of the nucleus size, $r_N\sim A^{1/3}(1\,\text{fm})\,$.
By requiring $r_N \sqrt{|t_N|_{\rm min} }\lesssim 1 $, 
we can estimate the maximal mass of a coherently-produced ALP: 
\begin{align}
    (m_a)_{\text{\tiny max} }
    \sim
    20\,\GeV
   \left( \frac{E_e}{18\,\GeV}\right)^{1/2}\left( \frac{E_N/A}{100\,\GeV} \right)^{1/2} \left(\frac{207}{A}\right)^{1/6}  \, .
   \label{eq:mamax}
\end{align}
In the left panel of Fig.~\ref{fig:diff_and_boost}, we show the normalized differential cross section $\mathrm{d}\sigma/\mathrm{d}\log |t|$ for $t=t_e$\,(solid) and $t=t_N$\,(dashed) for $m_a=\{0.1,1,10\}\,$GeV. 
As expected, the distributions for lighter ALP masses are centered around lower values of $\abs{t_e}$ and $\abs{t_N}$. 
The sharp drop in the ion momentum-transfer distribution is due to finite nucleon size.  
The inherent asymmetry of the production mechanism is reflected in the properties of the produced ALPs, which tend to be boosted in the same direction as the electron, see right panel of Fig.~\ref{fig:diff_and_boost} where we show $70\,\%$ contours of the 2D probability distribution in the $\{E_a/m_a,\eta_a\}$ plane. 
These contours are constructed by summing up the bins starting from the highest probability bin, until $70\,\%$ of the events are contained.
The production of heavier ALPs requires larger momentum extraction from the electron, while larger momentum extraction from the nucleus is suppressed by the form factor. 
Therefore, heavier ALPs are produced with a smaller boost and are more aligned with the electron.

\subsection{Prompt signal}
\label{sec: prompt}

\begin{figure}[t]
\centering
\includegraphics[width=0.32\textwidth]{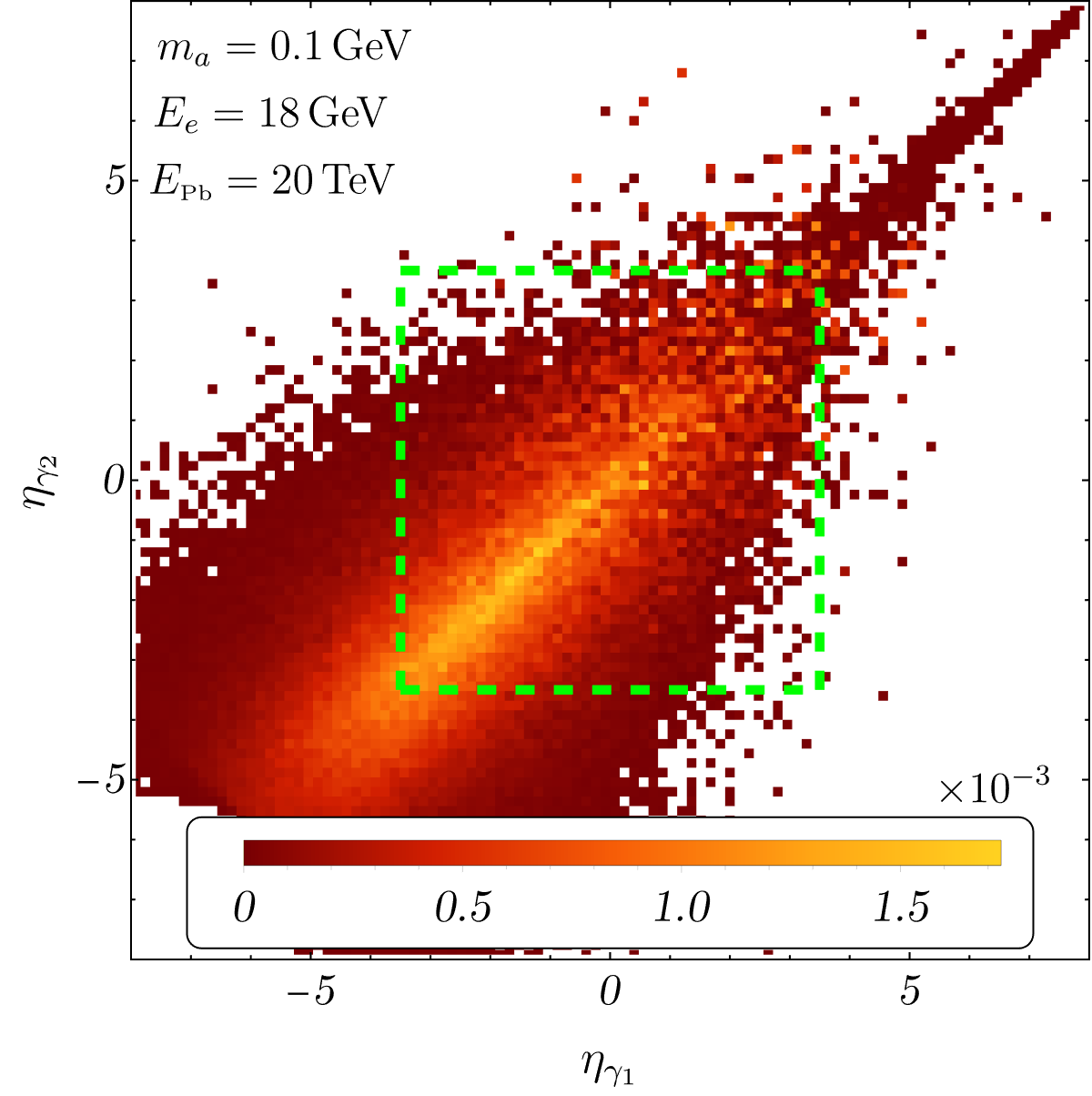}
\includegraphics[width=0.32\textwidth]{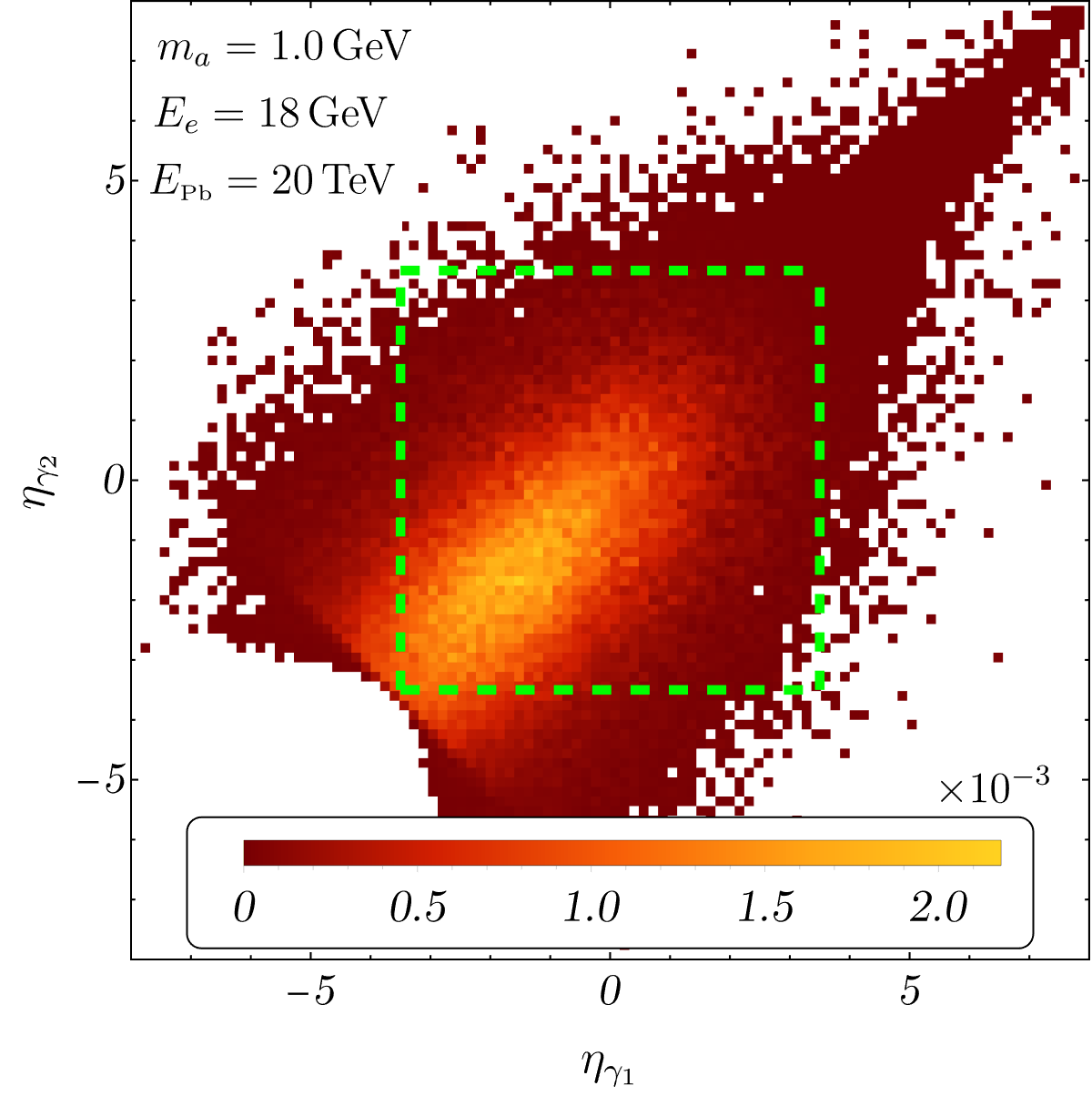}
\includegraphics[width=0.32\textwidth]{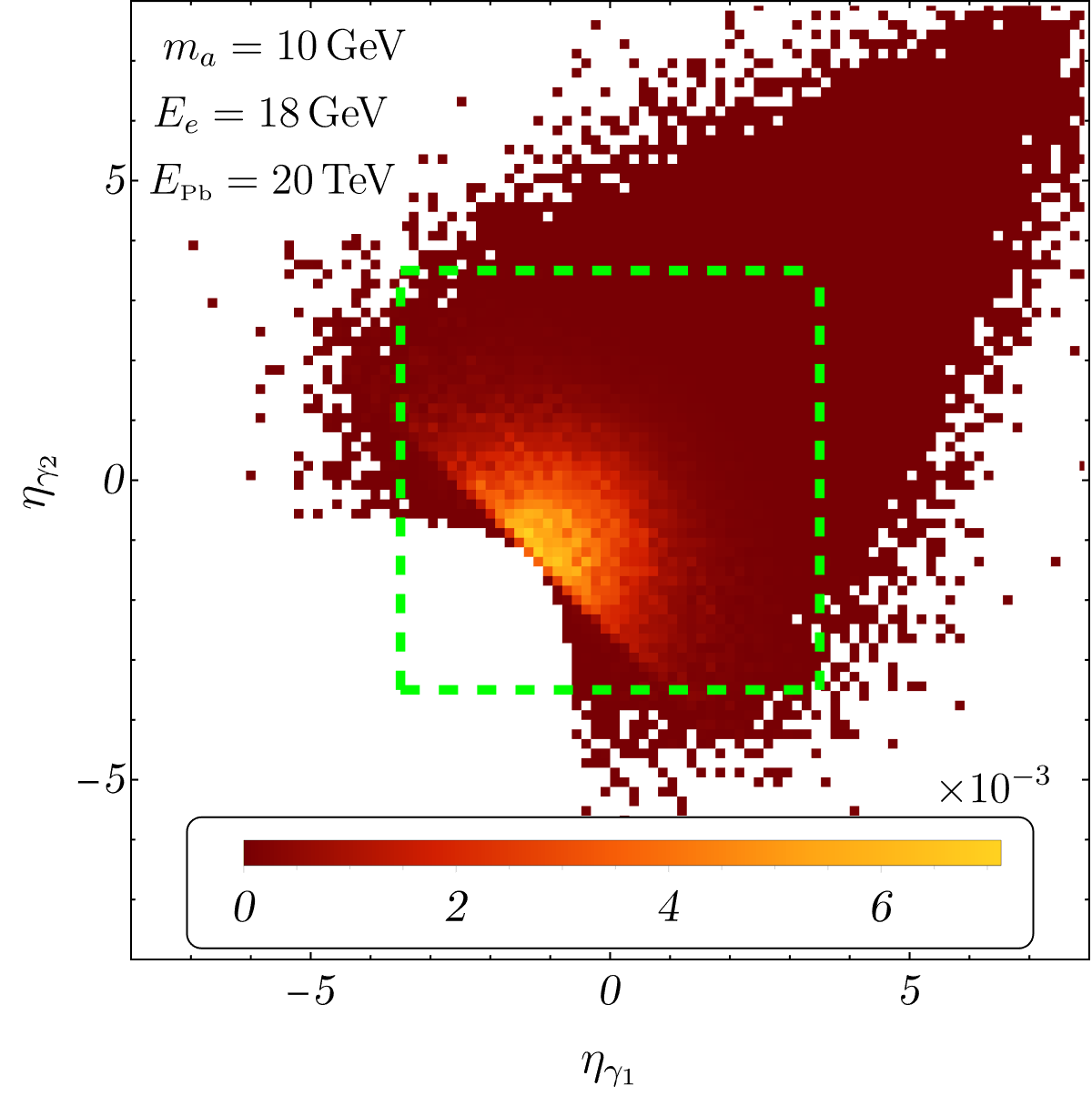}
	\caption{Pseudorapidity probability distribution for the two photons produced from an ALP decay for $m_a=0.1\,\GeV$ (left), $m_a=1.0\,\GeV$ (center), and $m_a=10\,\GeV$ (right). The central detector acceptance is indicated with the green dashed line square ( $-3.5< \eta < 3.5$). 
 }   
	\label{fig:eta}
\end{figure}

After being produced, the ALP decays back to photons with the rate $\Gamma_{a\to\gamma\gamma}= {m_a^3}/({64\pi \Lambda^2})$, which we assume is its main decay channel, namely $\BR(a\to \gamma \gamma) \approx 100\%$. 
The pseudorapidity distributions for the two final photons are presented in Fig.~\ref{fig:eta} for $m_a=\{0.1,1,10\}\,\GeV$.
As expected, the final state photons inherit the properties of the produced ALPs, discussed above.

As a result, the angular distributions of the photons are generally biased towards the negative $z$ direction, \ie{} negative $\eta$ values. 
Photons produced from heavier ALP decays are more likely to propagate in the negative $z$ direction and at larger angles (\ie{} smaller values of $|\eta|$) due to the fact that a heavier ALP is less boosted in the lab frame.

Our search strategy is to select events which, in addition to the recoiled electron, include two photons in the final state. 
The four-momentum of the recoiled ion cannot be directly measured at the EIC due to its extremely small scattering angles.
Since $\Gamma_{a\to\gamma \gamma} / m_a\ll1$, our signal would ideally appear as a narrow peak in the spectrum of the observed diphoton mass $\mgg$, defined in terms of the photon energies $E_1$, $E_2$ and their relative angle $\theta_{12}$,
\begin{align}
    \mgg = \sqrt{2E_{1}E_{2} (1-\cos\theta_{12})}\,.
\end{align}
The method chosen to construct the ALP mass from the two-photon final state is similar to the one used to reconstruct the $\pi^0$ in Ref.~\cite{Bock:2022lwp}. 
The reconstructed mass resolution $\Dmgg$ is determined by fitting the $m_{\gamma\gamma}$ spectrum, which includes the detector response and efficiency.
See Table~\ref{tab:mgg_res} for the values used in this work.

\begin{table}[b]
\centering
\begin{tabular}{|c|c|c|c|c|c|c|c|c|c|} 
 \hline
 $\mgg\,$[GeV]  & 0.3 & 0.5 & 0.7 & 0.9 & 2.0 &4.0 &7.0 & 15.0 \\
 \hline
$\Dmgg/\mgg$~(\%)  &  3.5 &  3.3 &  3.1 &  2.8 &  1.7 &  1.2 &  0.97 &  0.72 \\ 
 \hline
\end{tabular}
\caption{The reconstructed ALP mass resolution $\Dmgg/\mgg$ for selected $\mgg$ values, determined by fitting the $m_{\gamma\gamma}$ spectrum and including the detector response and efficiency.
}
\label{tab:mgg_res}
\end{table}

\begin{figure}[t]
\centering
\includegraphics[width=0.44\textwidth]{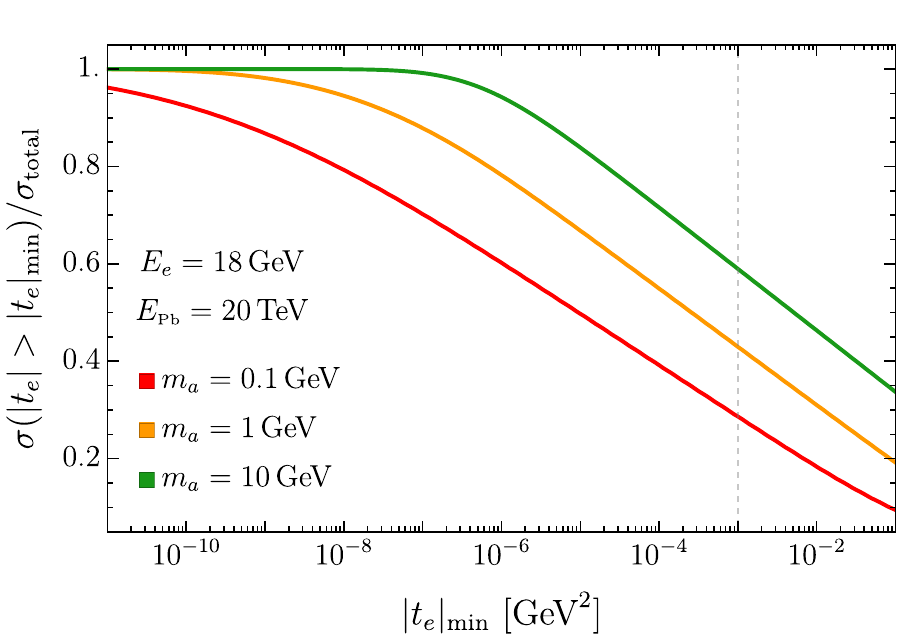}~~~~
\includegraphics[width=0.44\textwidth]{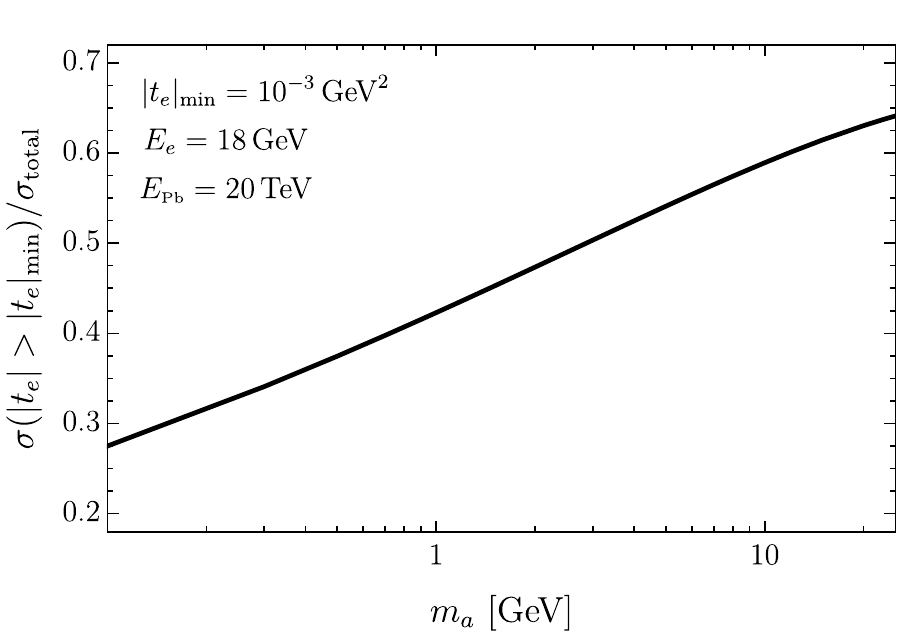}
 \caption{ Left panel: The $|t_e|$ cut signal efficiency as a function of $|t_e|_{\text{min}}$ for $m_a=\{0.1,1,10\}\,\GeV$ in red, orange and green, respectively. In dashed gray we mark the $|t_e|_{\text{min}}=10^{-3}\,\GeV^2$ used in this work. Right panel: Signal efficiency of the $|t_e|$ requirement shown in the left panel as a function of the ALP mass for $|t_e|_{\text{min}}=10^{-3}\,\GeV^2$.}	\label{fig:sigeff}
\end{figure}

The acceptance region is determined by the effective coverage of the calorimeters, see Table ~\ref{tab:energyresolution}. 
We ensure the photons are central enough to be detected by requiring that
\begin{align}
    \label{eq::acc}    
    \abs{\eta_{\gamma_{1}}}, \ \abs{\eta_{\gamma_{2}}} < 3.5 \,.
\end{align}
In addition, we place a requirement on the minimal photon energy,
\begin{align}
    \label{eq::basic_cut1}
    E_{\gamma_{1}},\ E_{\gamma_{2}} > 1.0\,\GeV\,,
\end{align}
which ensures the photons are energetic enough to be reliably reconstructed in the calorimeters and suppresses beam-induced backgrounds (such as radiative processes). Throughout this work, we conservatively assume a single photon efficiency detection of $\varepsilon_{\gamma}(E_\gamma,\eta_\gamma)\approx70\,\%$ which is  $E_\gamma$ and $\eta_\gamma$ independent.  

We also require a minimal electron momentum transfer,
\begin{align}
    \label{eq::basic_cut2}
    |t_e|>10^{-3}\,\GeV^2 \,,
\end{align}
so that the 4-momentum of the recoiled electron can be reconstructed reliably, see Sec.~11.7.2 of Ref.~\cite{AbdulKhalek:2021gbh}.
To demonstrate the effect of this cut on the signal efficiency, 
we show in the left panel of Fig.~\ref{fig:sigeff} the signal efficiency as a function of $|t_e|_{\text{min}}$ for $m_a=\{0.1,\,1,\,10\}\,\GeV$ and on the right panel as a function of mass for the value used in this work, $|t_e|_{\text{min}}=10^{-3}\,\GeV^2$. 

One potential source of signal reduction is due to photon merger, namely the inability to separate highly colinear photons appearing as a single merged object in the calorimeter. 
The EIC detectors are expected to be able to resolve two photons from the decay of neutral pions with $E_{\pi^0}\lesssim 10\,$GeV, see Sec.~11.3.2 of Ref.~\cite{AbdulKhalek:2021gbh}. 
More than $90\%$ of the lightest ALPs considered in this work, with $m_a \sim m_{\pi^0} \sim 0.1\,\GeV$, are produced with $E_a \leq 3\,\GeV$. 
Heavier ALPs are produced with even smaller average boost factors, see Fig.~\ref{fig:diff_and_boost}. 
Therefore, we estimate that photon merger should only have a negligible effect on the signal efficiency. 

\subsection{Prompt backgrounds}
\label{sec:background}

We consider three main sources of background due to SM processes: 
(1)~diphoton production via light-by-light scattering, 
(2)~$\pi^0$ pair-production with missing photons, and 
(3)~$\omega$ production with missing photons. 
All the background rates are estimated using the EPA, which approximates the virtual photons as being on-shell,  
thus implicitly taking the limit $|t_e| \to 0$, see Appendix~\ref{sec:EPA} for more details.
We therefore overestimate the background by not implementing the minimal $|t_e|$ requirement of Eq.~(\ref{eq::basic_cut2}). 
The backgrounds at each $m_a$ are estimated by integrating over the invariant mass in the window of $\mgg\in [m_a-2\Dmgg,m_a+2\Dmgg]$. 

The main background is due to the irreducible diphoton production,
\begin{align}
    e\,N \to e\,N\,\gamma\, \gamma\,,
    \label{eq:LBL}
\end{align}
where we focus on the $\cO(\alpha^3)$ process in which two virtual photons scatter off of each other via one-loop box diagrams, also known as light-by-light~(LBL) scattering.
For simplicity, the fermions in the loop are taken to be massless and the leading order LBL cross section is adopted from~\cite{Bern:2001dg}.
We take into account all SM fermions except for the top quark, thus overestimating the LBL background at scales below the bottom quark mass. 
Note we do not consider beam-induced backgrounds, \eg{} $\cO(\alpha^2)$ double bremsstrahlung, which we assume can be made negligible by requiring sufficiently central, Eq.~\eqref{eq::acc}, and energetic, Eq.~\eqref{eq::basic_cut1}, final state photons. 
These beam-related backgrounds are not well understood yet at the EIC, and a detailed estimation of them is beyond the scope of this work. 
The EPA estimations for the total ALP production cross section and diphoton production via LBL scattering are given in Eqs.~\eqref{eq:EPA_sig} and~\eqref{eq:EPA_LBL}, respectively.
In the right panel of Fig.~\ref{fig:feyn_and_xsecs}, we show in red the LBL cross section as a function of the diphoton mass $\mgg$ after applying all the requirements discussed in this section.

The remaining SM processes we consider involve more than two photons in the final state. 
They contribute as background only when some of the photons are missed, either due to the finite detector size or photon reconstruction inefficiency.  
First, we consider neutral pion pair production,
\begin{align}
    e\,N \to e\,N\,\pi^0\, \pi^0 \;\;\;\;\; 
    (\pi^0\, \pi^0 \to \gamma \gamma \gamma \gamma)\,,  
\end{align}
where two out of four final state photons are not detected. 
Several processes contribute to the neutral pion pair production~\cite{Klusek-Gawenda:2013rtu}: the light meson resonance contributions dominate in the low $\mgg$ region, while in the high $\mgg$ region, we use the hand-bag model~\cite{Diehl:2009yi} to estimate the production rate. 
Based on $4\times 10^6$ generated events, we simulate the detector effects by selecting which photons are missed.
First, photons that do not satisfy the basic acceptance and energy threshold requirements of Eqs.~\eqref{eq::acc} and~\eqref{eq::basic_cut1} are labeled as missing.
We then select which of the remaining photons are detected according to our conservative assumption of a single photon detection efficiency of 70\,\%. 

Due to the missing momentum in these events, it is useful to consider the reconstructed nuclear mass, defined as
\begin{align}
    (m_N^2)^{\text{recon.}} \equiv (k_e+k_n-p_e-p_{\gamma_1}-p_{\gamma_2})^2 = (p_n+p_{\gamma_3}+p_{\gamma_4})^2>m_N^2\,,
\end{align}
where $\gamma_1,\gamma_2$\,($\gamma_3,\gamma_4$) are the observed\,(missed) photons. 
Clearly, for events in which all final state particles are detected, $(m_N^2)^{\text{recon.}} \approx m_N^2$ up to the resolution of the measurement, which we estimate to be $\sim10\%$.
Therefore, this background can be efficiently reduced by requiring 
\begin{align}
    \label{eq::cut_mn_recon}
    (m_N^2)^{\text{recon.}} < (1.1 m_N)^2\,,
\end{align}
as well as requiring that the photons are back-to-back in the transverse plane, 
\begin{align}
    \label{eq::cut_b2b}
    |\pi-\Delta\varphi_{\gamma_1\gamma_2}| < 0.2\,,
\end{align}
where $\Delta \varphi_{\gamma_1\gamma_2}\equiv |\varphi_{\gamma_1}-\varphi_{\gamma_2}|$, \ie{} the difference of azimuthal angle between the two detected photons. 
Both the LBL scattering background and the signal, which have similar kinematics, are essentially unaffected by the cuts of Eqs.~\eqref{eq::cut_mn_recon} and \eqref{eq::cut_b2b}. 
The effective cross section of pion-pair production is shown in purple in the right panel of Fig.~\ref{fig:feyn_and_xsecs}. It is the dominant background at $\mgg\sim 1$~GeV.

Next, we consider the background due to $\omega$ production,
\begin{align}
    \label{eq::omega_process}
    e\, N \to e\, N\, \omega\;\;\;\;\;( \omega \to \pi^0 \, \gamma \to \gamma\, \gamma \,\gamma)\,,
\end{align}
which is only relevant in the mass region $m_a \lesssim m_{\omega} \approx 0.78\,\GeV$. 
We calculate this background level using a partial EPA, namely by calculating the rate of $N\,\gamma \to N\, \omega$, where only the photon emitted from the electron is approximated as on-shell. 
The details on the $\omega$ photoproduction are given in~\cite{Ballam:1972eq}.  
We then implement the decay chain of the $\omega$ meson to recover the kinematical distributions of the final three photons.
Similar to $\pi^0$ pair production, we simulate the detector effects by enforcing Eqs.~\eqref{eq::acc} and~\eqref{eq::basic_cut1} and randomly selecting which of the remaining photons are missed.
This background can be efficiently removed by requiring that in this mass range, both observed photons propagate backward \ie{} in the same direction as the electron.
Therefore, for $m_a < m_\omega$ we also apply 
\begin{align}
    \label{eq:eta_cut}
    \eta_{\gamma_1},\,\eta_{\gamma_2}<0\,.
\end{align}
Finally, the effective cross section of $\omega$ production is shown by the green line in the right panel of Fig.~\ref{fig:feyn_and_xsecs}.

Lastly, we consider the background due to the misidentification of electron-positron pairs, produced in the photon fusion process $\gamma \, \gamma \to e^+ \, e^-$. 
Assuming a $1\%$ misidentification rate, we find this background is subdominant compared to the LBL. 
We assume beam-induced backgrounds of this type, \eg{} electron pair emitted from the electron, can be removed similar to the 
double bremsstrahlung case, see comment below Eq.~\eqref{eq:LBL}. 

\begin{figure}[t]
\centering
\includegraphics[width=0.95\textwidth]{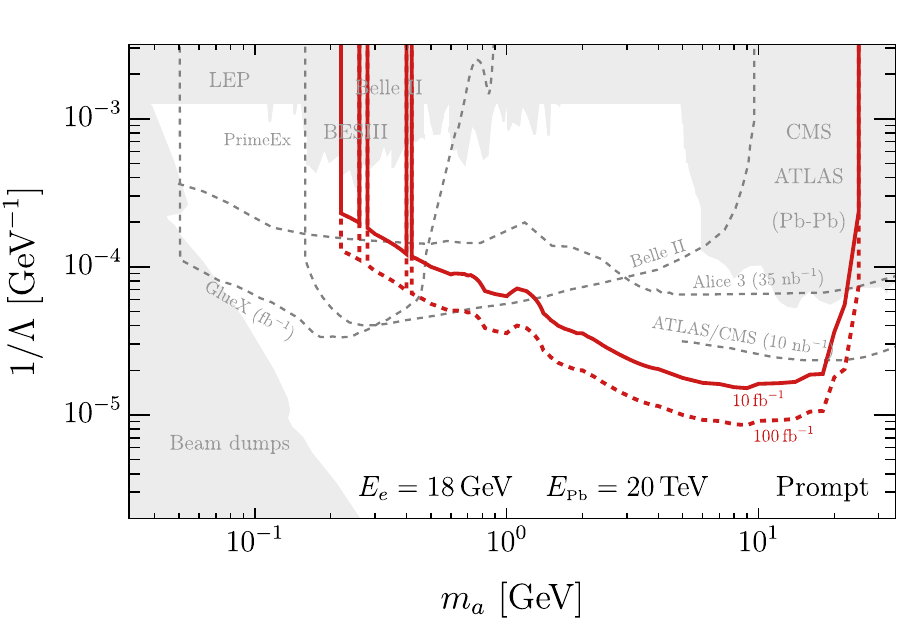}
\caption{The EIC projections on the prompt ALP searches with $E_e = 18\,$GeV and $E_{\text{Pb}} = 20\,$TeV. The solid (dashed) lines show the results with 10~(100) fb$^{-1}$ integrated luminosity.
In gray shaded regions are existing experimental constraints, while the dashed gray lines indicate projected sensitives for various proposed searches. }	
\label{fig:bound}
\end{figure}

We estimate the EIC sensitivity by requiring that $S_a = 2\sqrt{B}$, where $S_a$ is the number of ALP signal events and $B$ is the total number of background events in a window of $[m_a-2\Dmgg,m_a+2\Dmgg]$.
The projected sensitivities in the ALP parameter space using electron-lead collisions are shown as red curves in Fig.~\ref{fig:bound}, where the solid\,(dashed) lines correspond to the benchmark luminosity of 100\,(10)\,fb$^{-1}$.
Comparing the EIC to the current bounds from beam-dumps, LEP, Belle-II, BESIII and ATLAS/CMS~\cite{Bjorken:1988as,OPAL:2002vhf,Knapen:2016moh,Dobrich:2015jyk,Jaeckel:2015jla,Bauer:2017ris,Belle-II:2020jti,BESIII:2022rzz},
the EIC has the potential to probe unexplored parameter space at the heavy mass range $0.2 \lesssim m_a \lesssim 20\,\GeV$, reaching scales as high as $\Lambda \sim 10^5\,\GeV$ around $m_a = 10\,\GeV$.
Furthermore, the EIC can uniquely probe regions of the parameter space compared to other future experiments such as Belle-II~\cite{Dolan:2017osp}, heavy-ions~\cite{Knapen:2016moh,dEnterria:2022sut}, and Glue-X~\cite{Aloni:2019ruo}.

\subsection{Displaced-vertex search}
\label{sec:dvs}

The proposed BEMC design is hybrid, using light-collecting calorimetry and imaging calorimetry based silicon sensors~\cite{ATHENA:2022hxb}. 
The latter could provide information on the diphoton production vertex.
Therefore, the location of the ALP decay could be reconstructed and prompt SM backgrounds efficiently vetoed. 
This would allow the EIC to probe the parameter space of long-lived ALPs, namely ALPs which, in the lab frame, propagate a mean distance $L_a$ larger than the spatial resolution of the diphoton production vertex reconstruction, $L_R$, and smaller than the distance to the EM calorimeter, $L_{\rm EM}$.
The ALP decay probability inside this volume is approximated by
\begin{align}
    \cP(L_R, L_{\rm EM}) 
    \approx 
    \exp\left(-\frac{L_{R}}{L_a}\right) - \exp\left(-\frac{L_{\rm EM}}{L_a}\right) \, , 
\end{align}
where $L_a \equiv \beta\gamma/ \Gamma_a$.

\begin{figure}[t]
\centering
\includegraphics[width=0.49\textwidth]{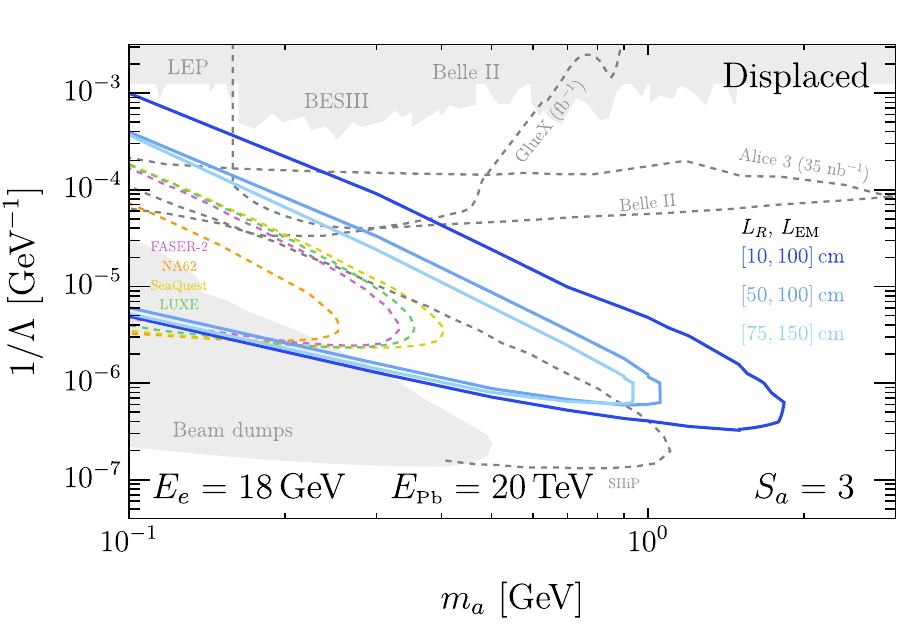}
\includegraphics[width=0.49\textwidth]{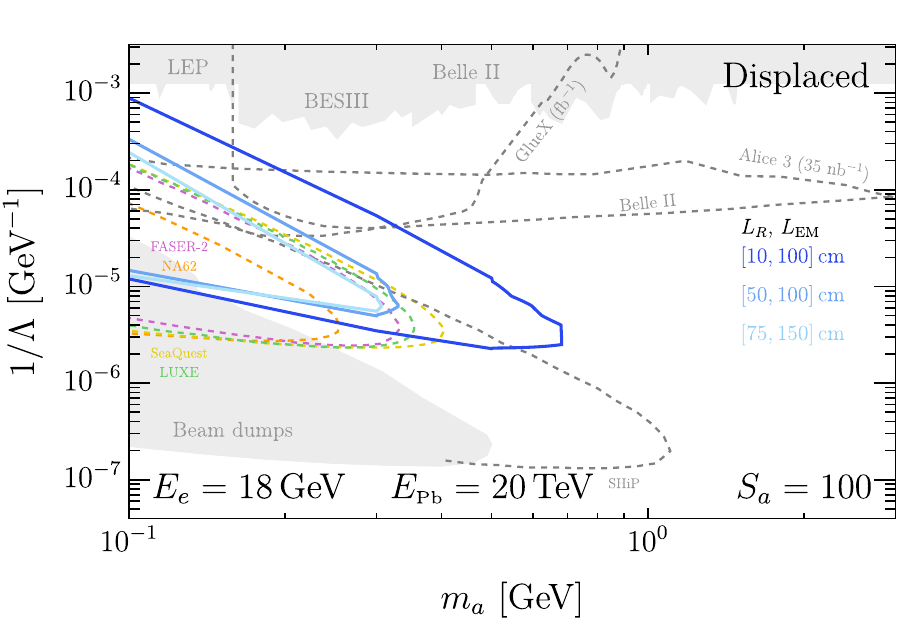}
\caption{The EIC projected sensitivity from a displaced ALP search with $E_e = 18$~GeV and $E_{\text{Pb}} = 20$\,TeV and $\cL=100\,\fb^{-1}$ defined by $S_a=3$ (\ie{} background free) and $S_a=100$ (\ie{} 2500 background events) on the left and right panel, respectively. 
The different shades of blue assume different diphoton spatial resolution, $L_R$, and distance between the interaction point and the EM calorimeter, $L_{\rm EM}$, as indicated on the plot.}	
\label{fig:dispalce}
\end{figure}

Since the EIC detector is still being designed, a full detector simulation with directional calorimetry is not yet available.
A detailed analysis of the future reach of the proposed displaced-vertex ALP search could only be done once the detector specifications are known. 
Instead, our goal is to show the potential of such a search strategy under reasonable assumptions regarding the vertex resolution of the BEMC.
For simplicity, we consider the following benchmarks: $[L_R, L_{\rm EM}]=[10,100]\,$cm, $[50,100]\,$cm and $[75,150]\,$cm. 
For background, we consider two cases: 
(1)~negligible background and 
(2)~$2500\left( \cL/100\,\fb^{-1} \right)$ background events per mass bin, such that $S_a=3$ and $S_a=100$ define our sensitivity curves, respectively.
Note that for this analysis we apply the minimal photon energy and pseudorapidity requirements of Eqs.~\eqref{eq::acc} and~\eqref{eq::basic_cut1}, as well as assuming a single photon detection efficiency of $70\%$. 
We do not apply the other requirements in Eqs.~\eqref{eq::basic_cut2}, \eqref{eq::cut_mn_recon}, \eqref{eq::cut_b2b}, and~\eqref{eq:eta_cut}.
The projections for the six benchmark combinations are presented in Fig.~\ref{fig:dispalce}.
As expected, a better spatial resolution, \ie{} smaller $L_R$, leads to increased sensitivity at shorter lifetimes, which is a blind spot for most other proposed experiments. 
We find that the displaced search has the potential to probe unexplored parameter space comparable to SHiP~\cite{Dobrich:2015jyk}.
Projections of displaced searches in NA62~\cite{Dobrich:2015jyk,Dobrich:2019dxc}, FASER~2~\cite{Feng:2018pew}, SeaQuest~\cite{Berlin:2018pwi} and LUXE~\cite{Bai:2021gbm}, which are expected to produce results before the EIC, are also shown in Fig.~\ref{fig:dispalce}.

\section{Outlook}
\label{sec:outlook}

The EIC is one of the major experiments planned for the next decade. 
Besides its vast SM program, it can be utilized to search for physics beyond the SM. 
In this work, we explore the case of ALPs, which are ubiquitous in various well-motivated BSM theories, coupled predominantly to photons.
In particular, we study the case of coherent ALP production via photon fusion, which enjoys a $Z^2$ enhancement and can support ALP masses of up to $\cO(20\,\GeV)$ due to the large boost of the ion relative to the electron.
Moreover, since the ions stay intact, the background can be efficiently vetoed and the search can be done in a relatively clean environment.

\begin{figure}[t]
\centering
\includegraphics[width=0.95\textwidth]{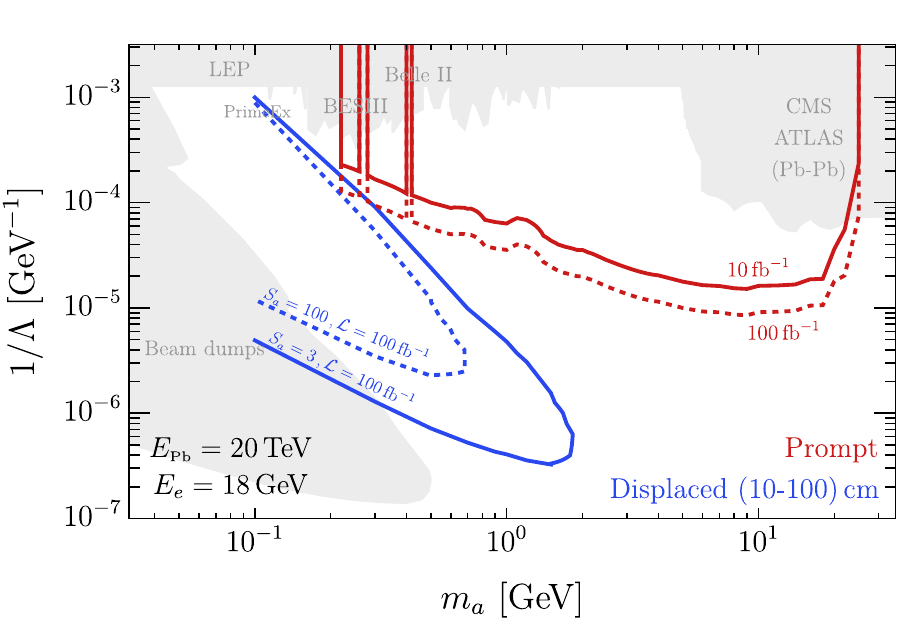}
\caption{
The EIC projections from the ALP searches with $E_e = 18\,$GeV and $E_{\text{Pb}} = 20\,$TeV. 
The solid (dashed) red lines show the prompt search results with 10~(100) fb$^{-1}$ integrated luminosity.
The solid (dashed) blue lines show the displaced search results with $S_a=3$~($S_a=100$) with 100 fb$^{-1}$ integrated luminosity, assuming the diphoton spatial resolution $L_R=10\,$cm and the distance between the interaction point and the EM calorimeter $L_{\rm EM}=100\,$cm.}	
\label{fig:promptdispalce}
\end{figure}

The predicted sensitivity for promptly decaying ALPs extends beyond current experimental collider limits, as well as future searches.
We find that the EIC can uniquely probe ALPs with a mass of $m_a\sim\cO(10\,\GeV)$ and effective scales as high as $\Lambda\sim 10^5\,\GeV$, see Fig~\ref{fig:bound}.
In addition, we briefly explore the case of long-lived ALPs which would appear as displaced vertices of photon pairs.
This type of search could be possible with sufficient vertex reconstruction resolution in the central EM calorimeter. 
Assuming spatial resolution of 10\,cm, ALPs as heavy as $\sim 1-2\,\GeV$ with effective scales as high as $\Lambda\sim \textrm{few} \times 10^6\,\GeV$ can be probed at the EIC, as shown in Fig~\ref{fig:dispalce}.
This study illustrates the importance of photon vertex reconstruction at the EIC in the context of BSM searches.
To conclude, we find that the EIC, under reasonable assumptions, would be in a unique position to probe large unexplored regions of the ALP-photon-coupling parameter space, as summarized in Fig~\ref{fig:promptdispalce}.

\acknowledgments
We thank Christoph Paus and Maria Zurek for useful discussions. 
R.B., H.L. and Y.S. are supported by grants from the ISF (grant No.~483/20), the BSF (grant No.~2020300), and by the Azrieli foundation.
T.M is partly supported by the research grants 2021-SGR-00649 and
PID2020-115845GB-I00/AEI/10.13039/501100011033 and Yan-Gui Talent Introduction Program (grant No.~118900M128). 
M.W.\ is supported by NSF grant PHY-2209181.
Y.S. and M.W. are supported by grant from the NSF-BSF (grant No.~2021800). W.L. is supported by Department of Energy grant DE-1020FG02-05ER41372.

\begin{appendices}

\section{ALP production cross section}
\label{sec:xs}

In this appendix, we give additional details about the $e^- N \to e^- N a$ calculation at the EIC including the EPA. 

\subsection{Squared amplitude of ALP production}

The 2-to-3 $e^- N \to e^- N a$ scattering amplitude can be written as 
\beq
    \cM^{2\to 3}_a 
=   \frac{e^2}{\Lambda}\frac{1}{t_N t_e}  
    J_{{\rm l},\mu} J^\mu_{\rm h} \, ,
\eeq
where $e$ is the EM gauge coupling, $J_{\rm l}^\mu$ and $J_{\rm h}^\mu$ are the leptonic and hadronic currents, respectively. 
The leptonic current is given by
\begin{align}
    J_{{\rm l}, \mu} 
=   \epsilon_{\mu\nu\alpha\beta} \bar{u}(p_e) \gamma^\nu u(k_e) 
    (k_n - p_n)^\alpha (p_e - k_e)^\beta\,.
\end{align}
The hadronic current is given by
\begin{align}
    J_{{\rm h},\mu} 
=   Z\,F(t_n)\,\left( k_n+p_n\right)_\mu \, ,
\end{align}
where $Z$ is the atomic number. 
We use the Helm form factor~\cite{Helm:1956zz}
\begin{align}
    F(q) 
=   \frac{3j_1(q R_1)}{q R_1}
    \text{exp}\left[-\frac{(q \rho)^2}{2}\right] \, ,
\end{align}
where $j_1$ is the first spherical Bessel function of the first kind, $\rho=0.9\,\fm = (0.22\,\GeV)^{-1}$ and 
\begin{align}
    R_1 
=   \sqrt{(1.23A^{1/3}-0.6)^2\,\text{fm}^2+({7}/{3})\pi^2 
    (0.52\,\fm)^2-5s^2} \approx 6.85\,\fm \approx (0.03\,\text{GeV})^{-1} \, ,
\end{align}
for a lead nucleus, $A=208$.

The spin-averaged squared amplitude is 
\begin{align}
    \abs{\cM^{2\to 3}_a}^2 
=   \frac{e^4}{\Lambda^2 t_N^2 t_e^2} L_{\mu\nu} W^{\mu\nu},
\end{align}
where the leptonic and hadronic tensors are defined as 
\begin{align}
    L_{\mu\nu} 
    \equiv 
    \frac{1}{2}\sum_{\text{spin}} J_{{\rm l}\mu}J_{{\rm l}\nu}^\dagger
    \quad \quad \text{and} \quad \quad
    W_{\mu\nu} 
    \equiv 
    J_{{\rm h}\mu}J_{{\rm h}\nu}^\dagger \, .
\end{align}

\subsection{Equivalent Photon Approximation}
\label{sec:EPA}

In the equivalent photon approximation, the differential cross section of a photon fusion process can be factorized as~\cite{Budnev:1975poe}
\begin{align}
    \label{eq:EPA}
    \frac{d\sigma_{eN\to eN X} }{d\hat{s}} (\hat{s})
=   \frac{1}{\hat{s}}\int_{\frac{\hat{s}}{4 E_{pb}}}^{E_e} \frac{d\omega_1}{\omega_1} 
    f_{\gamma/e}\left(\omega_1\right) f_{\gamma/N}\left(\omega_2\right)
    \hat{\sigma}_{\gamma\gamma\to X}(\hat{s})\, ,
\end{align}
where $\omega_1$\,($\omega_2$) is the photon energy emitted from the initial electron\,(ion),  $\hat{s} = 4 \omega_1\omega_2$ is the center-of-mass energy of the hard process and $\hat{\sigma}_{\gamma\gamma\to X}(\hat{s})$ is the $\gamma\gamma\to X$ cross section. 
The photon spectrum from the electron is
\begin{align}
    f_{\gamma/e}(\omega_1) 
=   \frac{\alpha}{2\pi}\left[1+\left(1-\frac{\omega_1}{E_e}\right)^2\right]
    \log\left(\frac{E_e^2}{m_e^2}\right) \, ,
\end{align}
where $\alpha \approx 1/137.036$ is the EM coupling constant. 
The photon spectrum from ion is taken from Ref.~\cite{Knapen:2016moh}
\begin{align}
    \label{eq:pdf}
    f_{\gamma/N}(\omega_2) 
=   \frac{2Z^2\alpha}{\pi}
    \left\{\frac{\omega_2}{E_R}K_0\left(\frac{\omega_2}{E_R}\right)K_1\left(\frac{\omega_2}{E_R}\right)
    \!-\!\frac{1}{2}\left(\frac{\omega_2}{E_R}\right)^2\left[K_1\left(\frac{\omega_2}{E_R}\right)^2\!-\!K_0\left(\frac{\omega_2}{E_R}\right)^2\right]\right\}  ,
\end{align}
where $E_R = E_{\rm Pb}/(M_{\rm Pb} \, R_A)$, $R_A = 1.2 A^{1/3}\,\fm$ and
$K_0$ and $K_1$ are the modified Bessel function of the second kind with order 0 and 1.

By using the narrow width approximation, we estimate the ALP production, $e^- N \to e^- N a$, to be 
\begin{align}
    \label{eq:EPA_sig}
    \sigma_{a}^{\rm EPA} 
=   \frac{\pi}{8 \Lambda^2} \int^{E_e}_{\frac{m_a^2}{4 E_{pb}}} 
    \frac{d\omega_1}{\omega_1}
    f_{\gamma/e}\left(\omega_1\right) f_{\gamma/N}\left(\frac{m_a^2}{4\omega_1}\right)\, .
\end{align}
We find that the ratio between the full 2-to-3 and the EPA cross sections is at most 2 in the relevant mass range. 

Considering all SM fermions, except the top, in the massless limit, the LBL scattering cross section at leading order $\hat{\sigma}^{\text{LO}}_{\text{LBL}}(\hat{s}) \simeq  10^{-6}/\hat{s}$~\cite{Bern:2001dg}. 
By using Eq.~\eqref{eq:EPA} with $\hat{\sigma}_{\gamma\gamma\to X}(\hat{s}) = \hat{\sigma}^{\text{LO}}_{\text{LBL}}(\hat{s})$,
the differential LBL cross section is given by
\begin{align}
    \label{eq:EPA_LBL}
    \frac{d \sigma^{\text{EPA}}_{\text{LBL}}}{d \mgg} 
    \simeq 
     \frac{2.0\times 10^{-6}}{\mgg^3} 
     \int^{E_e}_{\frac{\mgg}{4 E_{pb}}} 
     \frac{d\omega_1}{\omega_1}f_{\gamma/e}\left(\omega_1\right) 
     f_{\gamma/N}\left(\frac{\mgg^2}{4\omega_1}\right) \, .
\end{align}
For an ALP with mass $m_a$, the LBL background can be estimated by integrating over an interval of $[m_a-2\Dmgg,m_a+2\Dmgg]$.
Therefore, the ratio between the ALP signal and LBL background is roughly given by
\begin{align}
    \frac{S_a}{B_{\rm LBL}}
    \approx
    \frac{\sigma_a^{\rm EPA}}{4\Dmgg\frac{d \sigma^{\text{EPA}}_{\text{LBL}}}{d \mgg}}
    \approx
     20 \left( \frac{\TeV}{\Lambda}\right)^2 
     \left( \frac{\mgg}{2\,\GeV}\right)^2\left(\frac{0.01}{\Dmgg/\mgg} \right)\, .
\end{align}
%

\section{Dark photons at the EIC}
\label{sec:dark_photons}

In this appendix, we explore the possibility of searching for light dark photons~\cite{Okun:1982xi,Galison:1983pa,Holdom:1985ag,Pospelov:2007mp,Arkani-Hamed:2008hhe} at the EIC. 
We find that in the $(0.01-10)\,\GeV$ mass range, the dark photons are typically forward produced, collinearly with the electron. 
This makes the possibility of prompt searches challenging.
We estimate that using a prompt search strategy for a dimuon final state, the EIC can at best reach sensitives comparable to present bounds.
Alternatively, at sub-GeV masses, the dark photon may be long-lived enough such that it could be detected in an external decay volume.
However, the sensitivity of a displaced search strategy at the EIC depends crucially on the positioning of the decay volume. 
Due to the size and geometry of the EIC detector, we estimate the closest position for a decay volume at the EIC to be $\mathcal{O}(30\,\text{m})$ away from the interaction point.
This implies that the EIC sensitivity in a displaced search would not be able to improve on existing bounds.
For a preliminary estimation of the EIC reach for a dark photon with an effective propagation length of $\cO({\rm mm})$ see~\cite{Davoudiasl:2023pkq}.

This appendix is organized as follows. 
In Sec.~\ref{sec:dp_production}, we present the model and discuss the coherent production of dark photons at the EIC, with further details provided in Sec.~\ref{app:dark_photon_details}. 
In Sections~\ref{app:dark_photon_prompt} and~\ref{app:dark_photon_displaced} we briefly present our results for prompt and displaced searches, respectively.

\subsection{Coherent dark photon production}
\label{sec:dp_production}

\begin{figure}[t] 
\includegraphics[width=0.32\textwidth]{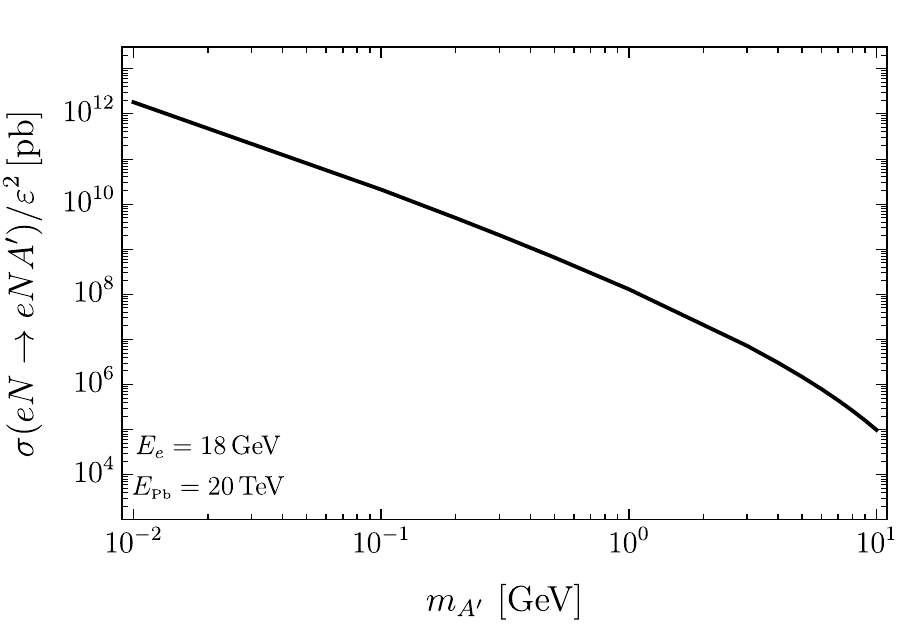}
\includegraphics[width=0.31\textwidth]{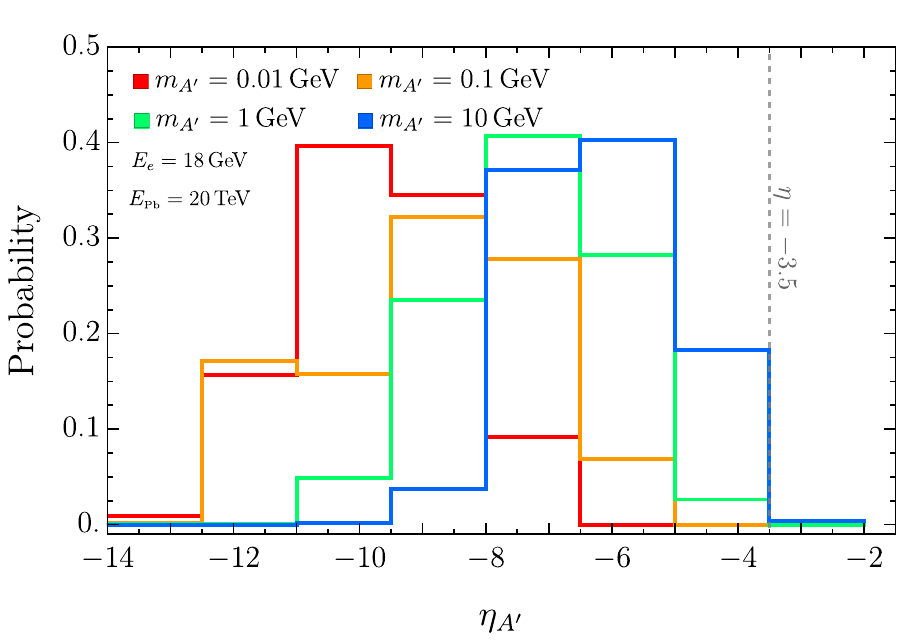}
\includegraphics[width=0.32
\textwidth]{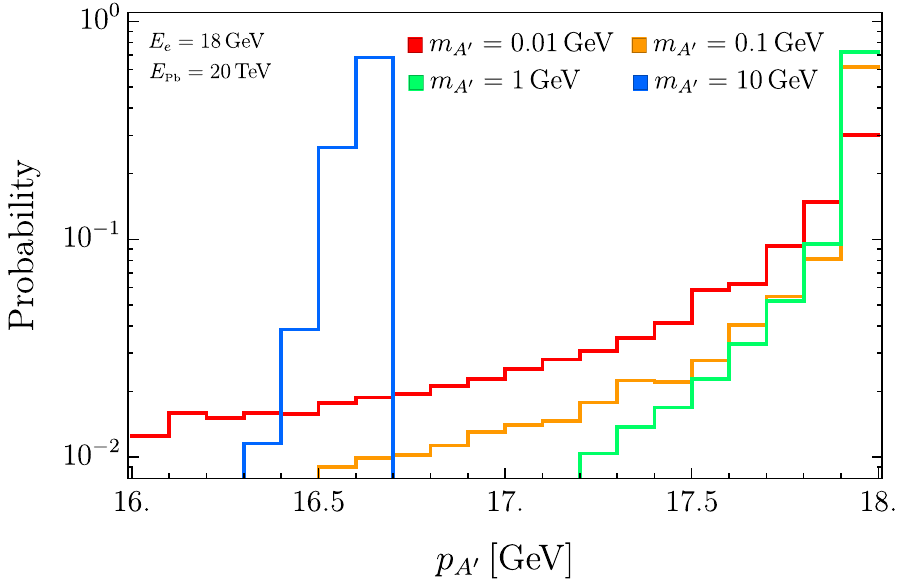}
	\caption{
	Left panel: The total dark photon production cross section. Right and middle panels: Dark photon pseudorapidity (middle) and lab-frame momentum (right) probability distributions for $m_{A'}=0.01,0.1,1 \text{ and } 10~$GeV in red, orange, green and blue, respectively.}	
	\label{fig:DP_eta_and_P_dist}
\end{figure}

The minimal dark photon model depends on two free parameters, namely the dark photon mass $m_{A'}$ and the mixing parameter $\varepsilon$, as defined by the Lagrangian
\begin{align}
    \mathcal{L}_{A'} = -\frac14 F^{\mu\nu}F_{\mu\nu}-\frac14 F^{'\mu\nu}F'_{\mu\nu}-\frac{\varepsilon}{2}F^{\mu\nu}F'_{\mu\nu}+\frac12 m_{A'}^2 {A'}^2+e A_\mu J^\mu_{\text{\tiny em}}\,,
\end{align}
where $F^{(\prime)}_{\mu\nu} \equiv \partial_\mu A^{(\prime)}_\nu-\partial_\nu A^{(\prime)}_\mu$. 
After moving to the kinetic canonical and mass basis, the physical dark photon, which for simplicity of notation shall remain $A'_\mu$, couples to the electromagnetic current, $\mathcal{L}_{A'} \supset \varepsilon e A'_\mu J^\mu_{\text{\tiny em}}\,$.
Thus, the electron can radiate an on-shell dark photon via a dark bremsstrahlung process
\begin{align}
    \label{eq:dp_2to3}
    e(k_e)+ N(k_N) 
    \;\; \to \;\; e(p_e)+N(p_N)+A'(p_{A'})\,.
\end{align}
Similar to the ALP production, the differential cross section of $e^-N\to e^- N A'$ depends on four variables, as well as on the (fixed) center-of-mass energy $\sqrt{s}$. 
For more details on the calculation of the full $2\to3$ production cross section, see Sec.~\ref{app:dark_photon_details}.
The total production cross section is shown in the left panel of Fig~\ref{fig:DP_eta_and_P_dist}.
The maximal mass for the coherently-produced dark photon, given by
\begin{align}
    (m_{A'})_{\text{\tiny max} }
    \sim
    20\,\GeV
   \left( \frac{E_e}{18\,\GeV}\right)^{1/2}\left( \frac{E_N/A}{100\,\GeV} \right)^{1/2} \left(\frac{207}{A}\right)^{1/6}  \, ,
   \label{eq:mDPmax}
\end{align}
is found following a derivation identical to the ALP case, see Sec.~(\ref{sec:ALP_production}).
We summarize the kinematics of the produced dark photon in the middle and right panels of Fig.~\ref{fig:DP_eta_and_P_dist}, where we show the pseudorapidity (momentum) distribution in the middle (right) panel for $m_{A'}=\{0.01,0.1,1,10\}\,$GeV in red, orange, green and blue, respectively. 
The dark photon is typically produced along the same direction as the electron, as shown in the left panel of Fig.~\ref{fig:DP_eta_and_P_dist}. 
The produced dark photon takes most of the electron energy in the lab frame, with only a small amount of its energy extracted from the nucleus, as expected from a coherent process.
Thus, the energy distributions are peaked around $E_{A'} \sim 18\,\GeV$, and the corresponding momentum distributions are peaked around $p_{A'}\sim({E_{A'}^2-m_{A'}^2})^{1/2}$, as shown in the right panel of Fig.~\ref{fig:DP_eta_and_P_dist}.
The dark photon lifetime and branching ratios are taken from Ref.~\cite{Ilten:2016tkc}.

\subsection{Prompt search}
\label{app:dark_photon_prompt}

\begin{figure}[t]
\centering
\includegraphics[width=0.49\textwidth]{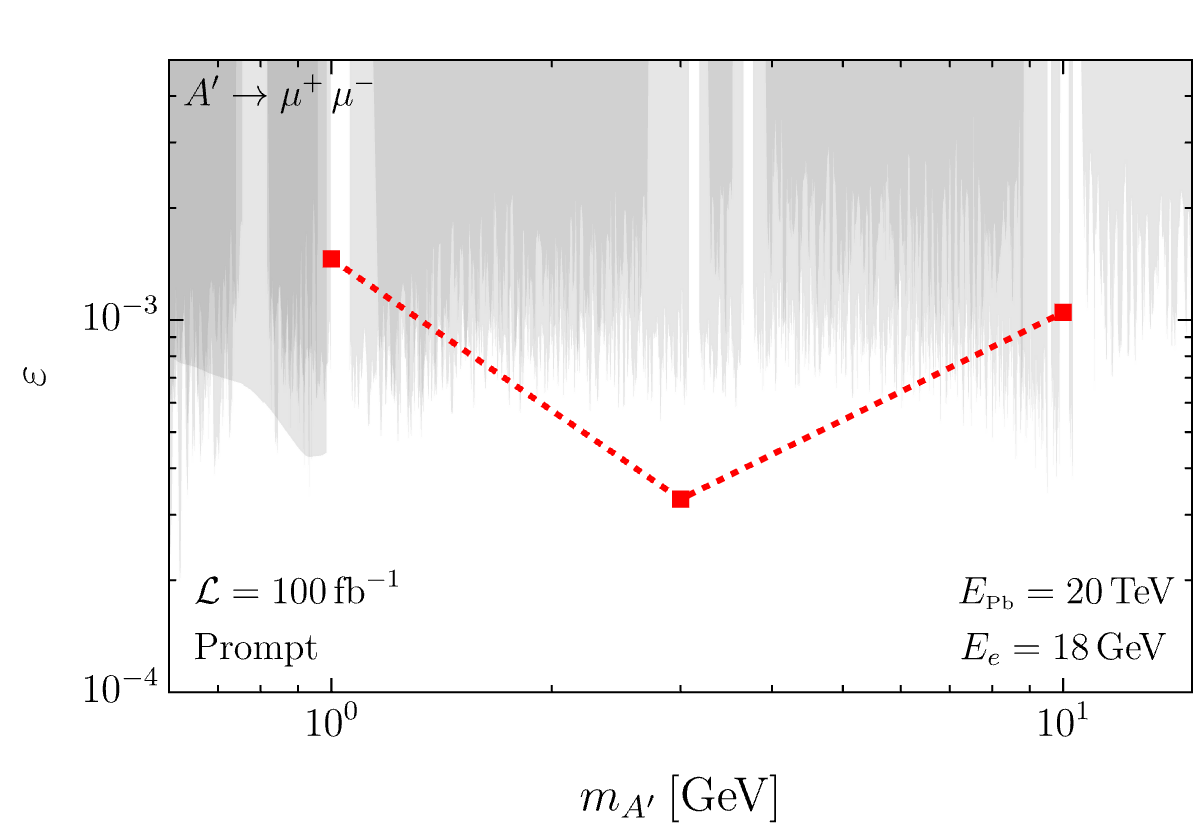}
\includegraphics[width=0.49\textwidth]{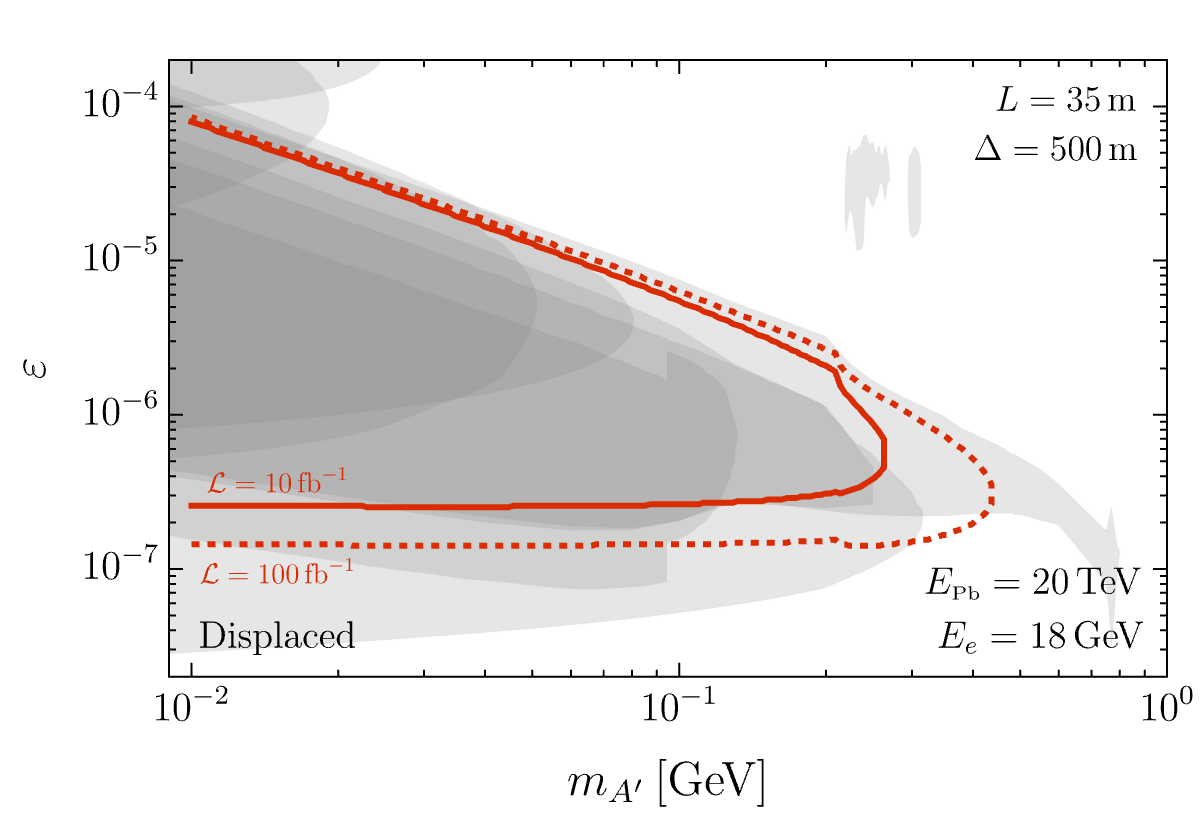}
\caption{Left panel: the projected sensitivity at the EIC for a prompt $A'\to \mu^+ \mu^-$ search with $\mathcal{L}=100\,\text{fb}^{-1}$ at $1,3$ and $10\,$GeV, interpolated with a dashed red line.
Right panel: the projected sensitivity at the EIC for a displaced $A'\to \mu^+ \mu^-$ search with $\mathcal{L}=10 $ and $100\,\text{fb}^{-1}$ in solid red and dashed red, respectively, assuming a $\Delta=500\,$m long, shielded decay volume located $L=35\,$m away from the interaction point.}	
\label{fig:sens_DP}
\end{figure}

We consider a prompt search in the mass range $1\,\GeV<m_{A'}<10\,\GeV$ with a dimuon final state $A'\to \mu^+ \mu^-$, chosen because of its sizable decay branching fraction and less background, including beam-induced background.
After calculating the $\mu^+\mu^-$ spectrum in the lab frame, we estimate the total number of events $S_{A'}$ which pass the acceptance criterion, namely
\begin{align}
    |\eta_{\mu^\pm}|<3.5\,,
\end{align}
and require that $S_{A'} = 2\sqrt{B_{\rm EM}}$, where $B_{\rm EM}$ is the irreducible EM-induced background estimated in~\cite{Ilten:2016tkc}.
We performed the analysis for three mass points $m_{A'} = 1,3 \text{ and }10\,\GeV$, enough to demonstrate the reach of the EIC for this search, shown in red in the left panel of Fig.~\ref{fig:sens_DP}. 
At $m_{A'}=1\,\GeV$, where the production rate is largest, the acceptance is $\mathcal{O}(10^{-5})$ due to the large boost of the produced dark photon, which reduces the sensitivity. 
On the other hand, acceptance at $m_{A'}=10\,\GeV$ is $\cO(1)$, but the sensitivity is suppressed due to the low coherent production rate.
In the middle of the considered mass range, we find the optimal EIC sensitivity, which is comparable to the existing bounds~\cite{Abrahamyan:2011gv,Lees:2014xha,Adrian:2018scb,Anastasi:2015qla,Anastasi:2018azp,Aaij:2019bvg,Batley:2015lha} shown in gray.

\subsection{Displaced search}
\label{app:dark_photon_displaced}

At lighter masses and smaller couplings, the dark photon decay is displaced from the interaction point. 
Due to the size and geometry of the EIC, we estimate that a $\Delta=500\,$m decay volume could be placed $L=35\,$m away from the interaction point.
The decay probability factor is then given by
\begin{align}
    P \approx e^{-L/L_{A'}}(1-e^{-\Delta/L_{A'}})\,,
\end{align}
assuming ideal acceptance, with $L_{A'} = \beta \gamma/\Gamma_{A'}$.
The resulting sensitivity contours defined by 3 signal events are shown in the right panel of Fig.~\ref{fig:sens_DP}, assuming ideal detection efficiency to all SM final states.
We further assume zero background events which can potentially be achieved with sufficient shielding. 
The EIC sensitivity, under the above assumptions 
does not improve on existing experimental bounds~\cite{Tsai:2019mtm,Andreas:2012mt,Riordan:1987aw,Konaka:1986cb,Banerjee:2019hmi,Astier:2001ck,Davier:1989wz,Bernardi:1985ny} shown in gray.

\subsection{Details on dark photon production}
\label{app:dark_photon_details}
 
The differential cross section for the $2\to3$ process of Eq.~\eqref{eq:dp_2to3} is given by
\begin{align}
    \frac{\mathrm{d}\sigma(eN \to eNA')}{\mathrm{d}m_{eA'}\,\mathrm{d}t\,\mathrm{d}\Omega_{A'}} 
&=  \frac{e^6 \varepsilon^2 p_{\text{\tiny cm}}(m_{eA'},m_{A'},m_e)}
    {(2\pi)^4 2^6p^2_{\text{\tiny cm}}(\sqrt{s},m_e,m_N) \, s}\, 
    \left(\frac{F(t)}{t} \right)^2\mathcal{A}\,.
\end{align}
where $p_{\text{\tiny cm}}(M,m_1,m_2)\equiv \sqrt{\lambda(M^2,m_1^2,m_2^2)}/(2M)$.
The four kinematic integration variables are $m^2_{eA'} \equiv (p_e+p_{A'})^2$, $t\equiv -(p_N-k_N)^2$,
and $\mathrm{d}\Omega_{A'} \equiv \mathrm{d}\cos\theta_{A'}\mathrm{d}\varphi_{A'}$, which define the directions of the dark photon momentum in the center-of-mass frame of the $e$-$A'$ system, \ie{} the system defined by $p_e+p_{A'} = (m_{eA'},0,0,0)$. The integration limits are 
\begin{align}
    -1 < \cos\theta_{A'} < 1\,, \quad
    0 < \varphi_{A'} < 2\pi\,, \quad
    m_{e}+m_{A'} < m_{eA'} < m_{eA'}^{\text{\tiny max}}(t)\,, \quad
    t_{-} < t < t_{+}\,,
\end{align}
with
\begin{align}
    m_{e A'}^{\text{\tiny max}}(t) 
    =\sqrt{s+m_N^2 \!-\! \frac{E_{\text{\tiny cm}}(\sqrt{s},m_N,m_e) \left(2m_N^2+t\right)- p_{\text{\tiny cm}}(\sqrt{s},m_N,m_e)\sqrt{t  \left(4m_N^2+t\right)}}{m_N^2/\sqrt{s}}}\,,
\end{align}
defining $E_{\text{\tiny cm}}(M,m_1,m_2) \equiv  {(M^2+m_1^2-m_2^2)}/{(2 M)}$. Finally, the kinematic integration limit for $t$ is given by
\begin{align}
    \frac{t_{\pm}}{2} 
    \equiv  
    &E_{\text{\tiny cm}}(\sqrt{s},m_N,m_e)E_{\text{\tiny cm}}(\sqrt{s},m_N,m_{e}+m_{\phi})
    \nonumber\\
    &\quad \pm p_{\text{\tiny cm}}(\sqrt{s},m_N,m_e)p_{\text{\tiny cm}}(\sqrt{s},m_N,m_{e}+m_{\phi})-m_N^2\,,
\end{align}
where $F(t)$ is the nuclear form factor, for which we use only the elastic one (see~\cite{Bjorken:2009mm} and references therein). 
Lastly, we have the amplitude~\cite{Liu:2017htz}
\begin{align}
    \mathcal{A} &= 
     2 \left(\frac{\tilde{s}^2+\tilde{u}^2}{\tilde{s}\tilde{u}}\right)(t+4m_N^2)
 -\frac{8t}{\tilde s \tilde u}(P\cdot k_e)^2 
 -\frac{8t}{\tilde s \tilde u}(P\cdot p_e)^2 \nonumber
\\
& -\frac{8t}{\tilde s \tilde u}\frac{t_2+m_{A'}^2}{2}(t+4m_N^2)
+ 2\frac{(\tilde{s}+\tilde{u})^2}{\tilde{s}^2\tilde{u}^2}(m_{A'}^2+2m_e^2) (t+4m_N^2) t
 \\
 & -8\frac{(\tilde{s}+\tilde{u})^2}{\tilde{s}^2\tilde{u}^2}(m_{A'}^2+2m_e^2) \left( \frac{\tilde{u} P\cdot k_e+\tilde{s} P\cdot p_e}{\tilde s+\tilde u}\right)\,, \nonumber
\end{align}
with
\begin{align}
    \tilde{s} \equiv 2p_e\cdot p_{A'}+m_{A'}^2\,, \quad
    \tilde{u} \equiv -2k_e\cdot p_{A'}+m_{A'}^2\, , \quad
    P \equiv p_N+k_N\,,\quad
    t_2 \equiv -2k_e\cdot p_e+2m_e^2\,. 
\end{align}
Note that, as opposed to~\cite{Liu:2017htz}, we use the mostly minus metric convention.

\end{appendices}


\bibliographystyle{JHEP}
\bibliography{ref}

\providecommand{\href}[2]{#2}\begingroup\raggedright\begin{thebibliography}{100}

\bibitem{Accardi:2012qut}
A.~Accardi et~al., {\it {Electron Ion Collider: The Next QCD Frontier}:
  {Understanding the glue that binds us all}},  {\em Eur. Phys. J. A} {\bf 52}
  (2016), no.~9 268, [\href{http://arxiv.org/abs/1212.1701}{{\tt
  arXiv:1212.1701}}].

\bibitem{AbdulKhalek:2021gbh}
R.~Abdul~Khalek et~al., {\it {Science Requirements and Detector Concepts for
  the Electron-Ion Collider: EIC Yellow Report}},
  \href{http://arxiv.org/abs/2103.05419}{{\tt arXiv:2103.05419}}.

\bibitem{Gonderinger:2010yn}
M.~Gonderinger and M.~J. Ramsey-Musolf, {\it {Electron-to-Tau Lepton Flavor
  Violation at the Electron-Ion Collider}},  {\em JHEP} {\bf 11} (2010) 045,
  [\href{http://arxiv.org/abs/1006.5063}{{\tt arXiv:1006.5063}}]. [Erratum:
  JHEP 05, 047 (2012)].

\bibitem{Cirigliano:2021img}
V.~Cirigliano, K.~Fuyuto, C.~Lee, E.~Mereghetti, and B.~Yan, {\it {Charged
  Lepton Flavor Violation at the EIC}},  {\em JHEP} {\bf 03} (2021) 256,
  [\href{http://arxiv.org/abs/2102.06176}{{\tt arXiv:2102.06176}}].

\bibitem{Davoudiasl:2021mjy}
H.~Davoudiasl, R.~Marcarelli, and E.~T. Neil, {\it {Lepton-Flavor-Violating
  ALPs at the Electron-Ion Collider: A Golden Opportunity}},
  \href{http://arxiv.org/abs/2112.04513}{{\tt arXiv:2112.04513}}.

\bibitem{Zhang:2022zuz}
J.~L. Zhang et~al., {\it {Search for e\textrightarrow{}\ensuremath{\tau}
  charged lepton flavor violation at the EIC with the ECCE detector}},  {\em
  Nucl. Instrum. Meth. A} {\bf 1053} (2023) 168276,
  [\href{http://arxiv.org/abs/2207.10261}{{\tt arXiv:2207.10261}}].

\bibitem{Batell:2022ogj}
B.~Batell, T.~Ghosh, T.~Han, and K.~Xie, {\it {Heavy Neutral Leptons at the
  Electron-Ion Collider}},  \href{http://arxiv.org/abs/2210.09287}{{\tt
  arXiv:2210.09287}}.

\bibitem{Yan:2022npz}
B.~Yan, {\it {Probing the dark photon via polarized DIS scattering at the HERA
  and EIC}},  {\em Phys. Lett. B} {\bf 833} (2022) 137384,
  [\href{http://arxiv.org/abs/2203.01510}{{\tt arXiv:2203.01510}}].

\bibitem{AbdulKhalek:2022hcn}
R.~Abdul~Khalek et~al., {\it {Snowmass 2021 White Paper: Electron Ion Collider
  for High Energy Physics}},  in {\em {2022 Snowmass Summer Study}}, 3, 2022.
\newblock \href{http://arxiv.org/abs/2203.13199}{{\tt arXiv:2203.13199}}.

\bibitem{Davoudiasl:2023pkq}
H.~Davoudiasl, R.~Marcarelli, and E.~T. Neil, {\it {Displaced Signals of Hidden
  Vectors at the Electron-Ion Collider}},
  \href{http://arxiv.org/abs/2307.00102}{{\tt arXiv:2307.00102}}.

\bibitem{Boughezal:2022pmb}
R.~Boughezal, A.~Emmert, T.~Kutz, S.~Mantry, M.~Nycz, F.~Petriello,
  K.~\c{S}im\c{s}ek, D.~Wiegand, and X.~Zheng, {\it {Neutral-current
  electroweak physics and SMEFT studies at the EIC}},  {\em Phys. Rev. D} {\bf
  106} (2022), no.~1 016006, [\href{http://arxiv.org/abs/2204.07557}{{\tt
  arXiv:2204.07557}}].

\bibitem{Liu:2021lan}
Y.~Liu and B.~Yan, {\it {Searching for the axion-like particle at the EIC}},
  \href{http://arxiv.org/abs/2112.02477}{{\tt arXiv:2112.02477}}.

\bibitem{Yue:2023mew}
C.-X. Yue, H.~Wang, X.-J. Cheng, and Y.-Q. Wang, {\it {Sensitivity of the
  future e-p collider to the coupling of axionlike particles with vector
  bosons}},  {\em Phys. Rev. D} {\bf 107} (2023), no.~11 115025,
  [\href{http://arxiv.org/abs/2305.19561}{{\tt arXiv:2305.19561}}].

\bibitem{Wang:2024zns}
H.-L. Wang, X.-K. Wen, H.~Xing, and B.~Yan, {\it {Probing the four-fermion
  operators via the transverse double spin asymmetry at the Electron-Ion
  Collider}},  \href{http://arxiv.org/abs/2401.08419}{{\tt arXiv:2401.08419}}.

\bibitem{Peccei:1977hh}
R.~D. Peccei and H.~R. Quinn, {\it {CP Conservation in the Presence of
  Instantons}},  {\em Phys. Rev. Lett.} {\bf 38} (1977) 1440--1443.

\bibitem{Peccei:1977ur}
R.~D. Peccei and H.~R. Quinn, {\it {Constraints Imposed by CP Conservation in
  the Presence of Instantons}},  {\em Phys. Rev. D} {\bf 16} (1977) 1791--1797.

\bibitem{Weinberg:1977ma}
S.~Weinberg, {\it {A New Light Boson?}},  {\em Phys. Rev. Lett.} {\bf 40}
  (1978) 223--226.

\bibitem{Wilczek:1977pj}
F.~Wilczek, {\it {Problem of Strong $P$ and $T$ Invariance in the Presence of
  Instantons}},  {\em Phys. Rev. Lett.} {\bf 40} (1978) 279--282.

\bibitem{Witten:1984dg}
E.~Witten, {\it {Some Properties of O(32) Superstrings}},  {\em Phys. Lett. B}
  {\bf 149} (1984) 351--356.

\bibitem{Svrcek:2006yi}
P.~Svrcek and E.~Witten, {\it {Axions In String Theory}},  {\em JHEP} {\bf 06}
  (2006) 051, [\href{http://arxiv.org/abs/hep-th/0605206}{{\tt
  hep-th/0605206}}].

\bibitem{Conlon:2006tq}
J.~P. Conlon, {\it {The QCD axion and moduli stabilisation}},  {\em JHEP} {\bf
  05} (2006) 078, [\href{http://arxiv.org/abs/hep-th/0602233}{{\tt
  hep-th/0602233}}].

\bibitem{Nomura:2008ru}
Y.~Nomura and J.~Thaler, {\it {Dark Matter through the Axion Portal}},  {\em
  Phys. Rev. D} {\bf 79} (2009) 075008,
  [\href{http://arxiv.org/abs/0810.5397}{{\tt arXiv:0810.5397}}].

\bibitem{Freytsis:2010ne}
M.~Freytsis and Z.~Ligeti, {\it {On dark matter models with uniquely
  spin-dependent detection possibilities}},  {\em Phys. Rev. D} {\bf 83} (2011)
  115009, [\href{http://arxiv.org/abs/1012.5317}{{\tt arXiv:1012.5317}}].

\bibitem{Dolan:2014ska}
M.~J. Dolan, F.~Kahlhoefer, C.~McCabe, and K.~Schmidt-Hoberg, {\it {A taste of
  dark matter: Flavour constraints on pseudoscalar mediators}},  {\em JHEP}
  {\bf 03} (2015) 171, [\href{http://arxiv.org/abs/1412.5174}{{\tt
  arXiv:1412.5174}}]. [Erratum: JHEP 07, 103 (2015)].

\bibitem{Hochberg:2018rjs}
Y.~Hochberg, E.~Kuflik, R.~Mcgehee, H.~Murayama, and K.~Schutz, {\it {Strongly
  interacting massive particles through the axion portal}},  {\em Phys. Rev. D}
  {\bf 98} (2018), no.~11 115031, [\href{http://arxiv.org/abs/1806.10139}{{\tt
  arXiv:1806.10139}}].

\bibitem{Ghosh:2023tyz}
D.~K. Ghosh, A.~Ghoshal, and S.~Jeesun, {\it {Axion-like particle (ALP) portal
  freeze-in dark matter confronting ALP search experiments}},
  \href{http://arxiv.org/abs/2305.09188}{{\tt arXiv:2305.09188}}.

\bibitem{Dror:2023fyd}
J.~Dror, S.~Gori, and P.~Munbodh, {\it {QCD Axion-Mediated Dark Matter}},
  \href{http://arxiv.org/abs/2306.03145}{{\tt arXiv:2306.03145}}.

\bibitem{Fitzpatrick:2023xks}
P.~J. Fitzpatrick, Y.~Hochberg, E.~Kuflik, R.~Ovadia, and Y.~Soreq, {\it {Dark
  Matter Through the Axion-Gluon Portal}},
  \href{http://arxiv.org/abs/2306.03128}{{\tt arXiv:2306.03128}}.

\bibitem{Preskill:1982cy}
J.~Preskill, M.~B. Wise, and F.~Wilczek, {\it {Cosmology of the Invisible
  Axion}},  {\em Phys. Lett. B} {\bf 120} (1983) 127--132.

\bibitem{Abbott:1982af}
L.~F. Abbott and P.~Sikivie, {\it {A Cosmological Bound on the Invisible
  Axion}},  {\em Phys. Lett. B} {\bf 120} (1983) 133--136.

\bibitem{Dine:1982ah}
M.~Dine and W.~Fischler, {\it {The Not So Harmless Axion}},  {\em Phys. Lett.
  B} {\bf 120} (1983) 137--141.

\bibitem{Aloni:2019ruo}
D.~Aloni, C.~Fanelli, Y.~Soreq, and M.~Williams, {\it {Photoproduction of
  Axionlike Particles}},  {\em Phys. Rev. Lett.} {\bf 123} (2019), no.~7
  071801, [\href{http://arxiv.org/abs/1903.03586}{{\tt arXiv:1903.03586}}].

\bibitem{CHARM:1985anb}
{\bf CHARM} Collaboration, F.~Bergsma et~al., {\it {Search for Axion Like
  Particle Production in 400-{GeV} Proton - Copper Interactions}},  {\em Phys.
  Lett. B} {\bf 157} (1985) 458--462.

\bibitem{Riordan:1987aw}
E.~M. Riordan et~al., {\it {A Search for Short Lived Axions in an Electron Beam
  Dump Experiment}},  {\em Phys. Rev. Lett.} {\bf 59} (1987) 755.

\bibitem{Bjorken:1988as}
J.~D. Bjorken, S.~Ecklund, W.~R. Nelson, A.~Abashian, C.~Church, B.~Lu, L.~W.
  Mo, T.~A. Nunamaker, and P.~Rassmann, {\it {Search for Neutral Metastable
  Penetrating Particles Produced in the SLAC Beam Dump}},  {\em Phys. Rev. D}
  {\bf 38} (1988) 3375.

\bibitem{Blumlein:1990ay}
J.~Blumlein et~al., {\it {Limits on neutral light scalar and pseudoscalar
  particles in a proton beam dump experiment}},  {\em Z. Phys. C} {\bf 51}
  (1991) 341--350.

\bibitem{Dobrich:2015jyk}
B.~D\"obrich, J.~Jaeckel, F.~Kahlhoefer, A.~Ringwald, and K.~Schmidt-Hoberg,
  {\it {ALPtraum: ALP production in proton beam dump experiments}},  {\em JHEP}
  {\bf 02} (2016) 018, [\href{http://arxiv.org/abs/1512.03069}{{\tt
  arXiv:1512.03069}}].

\bibitem{Dobrich:2017gcm}
B.~D\"obrich, {\it {Axion-like Particles from Primakov production in
  beam-dumps}},  {\em CERN Proc.} {\bf 1} (2018) 253,
  [\href{http://arxiv.org/abs/1708.05776}{{\tt arXiv:1708.05776}}].

\bibitem{Harland-Lang:2019zur}
L.~Harland-Lang, J.~Jaeckel, and M.~Spannowsky, {\it {A fresh look at ALP
  searches in fixed target experiments}},  {\em Phys. Lett. B} {\bf 793} (2019)
  281--289, [\href{http://arxiv.org/abs/1902.04878}{{\tt arXiv:1902.04878}}].

\bibitem{Dobrich:2019dxc}
B.~D\"obrich, J.~Jaeckel, and T.~Spadaro, {\it {Light in the beam dump - ALP
  production from decay photons in proton beam-dumps}},  {\em JHEP} {\bf 05}
  (2019) 213, [\href{http://arxiv.org/abs/1904.02091}{{\tt arXiv:1904.02091}}].
  [Erratum: JHEP 10, 046 (2020)].

\bibitem{NA64:2020qwq}
{\bf NA64} Collaboration, D.~Banerjee et~al., {\it {Search for Axionlike and
  Scalar Particles with the NA64 Experiment}},  {\em Phys. Rev. Lett.} {\bf
  125} (2020), no.~8 081801, [\href{http://arxiv.org/abs/2005.02710}{{\tt
  arXiv:2005.02710}}].

\bibitem{Afik:2023mhj}
Y.~Afik, B.~D\"obrich, J.~Jerhot, Y.~Soreq, and K.~Tobioka, {\it {Probing
  Long-lived Axions at the KOTO Experiment}},
  \href{http://arxiv.org/abs/2303.01521}{{\tt arXiv:2303.01521}}.

\bibitem{Ema:2023tjg}
Y.~Ema, Z.~Liu, and R.~Plestid, {\it {Searching for axions with kaon decay at
  rest}},  \href{http://arxiv.org/abs/2308.08589}{{\tt arXiv:2308.08589}}.

\bibitem{Belle-II:2020jti}
{\bf Belle-II} Collaboration, F.~Abudin\'en et~al., {\it {Search for Axion-Like
  Particles produced in $e^+e^-$ collisions at Belle II}},  {\em Phys. Rev.
  Lett.} {\bf 125} (2020), no.~16 161806,
  [\href{http://arxiv.org/abs/2007.13071}{{\tt arXiv:2007.13071}}].

\bibitem{Feng:2018pew}
J.~L. Feng, I.~Galon, F.~Kling, and S.~Trojanowski, {\it {Axionlike particles
  at FASER: The LHC as a photon beam dump}},  {\em Phys. Rev. D} {\bf 98}
  (2018), no.~5 055021, [\href{http://arxiv.org/abs/1806.02348}{{\tt
  arXiv:1806.02348}}].

\bibitem{Balkin:2021jdr}
R.~Balkin, M.~W. Krasny, T.~Ma, B.~R. Safdi, and Y.~Soreq, {\it {Probing ALPs
  at the CERN Gamma Factory}},  \href{http://arxiv.org/abs/2105.15072}{{\tt
  arXiv:2105.15072}}.

\bibitem{GlueX:2021myx}
{\bf GlueX} Collaboration, S.~Adhikari et~al., {\it {Search for photoproduction
  of axionlike particles at GlueX}},  {\em Phys. Rev. D} {\bf 105} (2022),
  no.~5 052007, [\href{http://arxiv.org/abs/2109.13439}{{\tt
  arXiv:2109.13439}}].

\bibitem{Pybus:2023yex}
J.~R. Pybus et~al., {\it {Search for axion-like particles through nuclear
  Primakoff production using the GlueX detector}},
  \href{http://arxiv.org/abs/2308.06339}{{\tt arXiv:2308.06339}}.

\bibitem{OPAL:2002vhf}
{\bf OPAL} Collaboration, G.~Abbiendi et~al., {\it {Multiphoton production in
  e+ e- collisions at s**(1/2) = 181-GeV to 209-GeV}},  {\em Eur. Phys. J. C}
  {\bf 26} (2003) 331--344, [\href{http://arxiv.org/abs/hep-ex/0210016}{{\tt
  hep-ex/0210016}}].

\bibitem{Jaeckel:2015jla}
J.~Jaeckel and M.~Spannowsky, {\it {Probing MeV to 90 GeV axion-like particles
  with LEP and LHC}},  {\em Phys. Lett. B} {\bf 753} (2016) 482--487,
  [\href{http://arxiv.org/abs/1509.00476}{{\tt arXiv:1509.00476}}].

\bibitem{Yue:2021iiu}
C.-X. Yue, H.-Y. Zhang, and H.~Wang, {\it {Production of axion-like particles
  via vector boson fusion at future electron-positron colliders}},  {\em Eur.
  Phys. J. C} {\bf 82} (2022), no.~1 88,
  [\href{http://arxiv.org/abs/2112.11604}{{\tt arXiv:2112.11604}}].

\bibitem{Tian:2022rsi}
M.~Tian, Z.~S. Wang, and K.~Wang, {\it {Search for long-lived axions with far
  detectors at future lepton colliders}},
  \href{http://arxiv.org/abs/2201.08960}{{\tt arXiv:2201.08960}}.

\bibitem{Bao:2022onq}
Y.~Bao, J.~Fan, and L.~Li, {\it {Electroweak ALP Searches at a Muon Collider}},
   \href{http://arxiv.org/abs/2203.04328}{{\tt arXiv:2203.04328}}.

\bibitem{BESIII:2022rzz}
{\bf BESIII} Collaboration, M.~Ablikim et~al., {\it {Search for an axion-like
  particle in radiative J/\ensuremath{\psi} decays}},  {\em Phys. Lett. B} {\bf
  838} (2023) 137698, [\href{http://arxiv.org/abs/2211.12699}{{\tt
  arXiv:2211.12699}}].

\bibitem{CMS:2012cve}
{\bf CMS} Collaboration, S.~Chatrchyan et~al., {\it {Search for Exclusive or
  Semi-Exclusive Photon Pair Production and Observation of Exclusive and
  Semi-Exclusive Electron Pair Production in $pp$ Collisions at $\sqrt{s}=7$
  TeV}},  {\em JHEP} {\bf 11} (2012) 080,
  [\href{http://arxiv.org/abs/1209.1666}{{\tt arXiv:1209.1666}}].

\bibitem{ATLAS:2014jdv}
{\bf ATLAS} Collaboration, G.~Aad et~al., {\it {Search for Scalar Diphoton
  Resonances in the Mass Range $65-600$ GeV with the ATLAS Detector in $pp$
  Collision Data at $\sqrt{s}$ = 8 $TeV$}},  {\em Phys. Rev. Lett.} {\bf 113}
  (2014), no.~17 171801, [\href{http://arxiv.org/abs/1407.6583}{{\tt
  arXiv:1407.6583}}].

\bibitem{Mimasu:2014nea}
K.~Mimasu and V.~Sanz, {\it {ALPs at Colliders}},  {\em JHEP} {\bf 06} (2015)
  173, [\href{http://arxiv.org/abs/1409.4792}{{\tt arXiv:1409.4792}}].

\bibitem{ATLAS:2015rsn}
{\bf ATLAS} Collaboration, G.~Aad et~al., {\it {Search for new phenomena in
  events with at least three photons collected in $pp$ collisions at $\sqrt{s}$
  = 8 TeV with the ATLAS detector}},  {\em Eur. Phys. J. C} {\bf 76} (2016),
  no.~4 210, [\href{http://arxiv.org/abs/1509.05051}{{\tt arXiv:1509.05051}}].

\bibitem{Brivio:2017ije}
I.~Brivio, M.~B. Gavela, L.~Merlo, K.~Mimasu, J.~M. No, R.~del Rey, and
  V.~Sanz, {\it {ALPs Effective Field Theory and Collider Signatures}},  {\em
  Eur. Phys. J. C} {\bf 77} (2017), no.~8 572,
  [\href{http://arxiv.org/abs/1701.05379}{{\tt arXiv:1701.05379}}].

\bibitem{Bauer:2017ris}
M.~Bauer, M.~Neubert, and A.~Thamm, {\it {Collider Probes of Axion-Like
  Particles}},  {\em JHEP} {\bf 12} (2017) 044,
  [\href{http://arxiv.org/abs/1708.00443}{{\tt arXiv:1708.00443}}].

\bibitem{Ebadi:2019gij}
J.~Ebadi, S.~Khatibi, and M.~Mohammadi~Najafabadi, {\it {New probes for
  axionlike particles at hadron colliders}},  {\em Phys. Rev. D} {\bf 100}
  (2019), no.~1 015016, [\href{http://arxiv.org/abs/1901.03061}{{\tt
  arXiv:1901.03061}}].

\bibitem{Bonilla:2022pxu}
J.~Bonilla, I.~Brivio, J.~Machado-Rodr\'\i{}guez, and J.~F. de~Troc\'oniz, {\it
  {Nonresonant searches for axion-like particles in vector boson scattering
  processes at the LHC}},  {\em JHEP} {\bf 06} (2022) 113,
  [\href{http://arxiv.org/abs/2202.03450}{{\tt arXiv:2202.03450}}].

\bibitem{Alonso-Alvarez:2023wni}
G.~Alonso-\'Alvarez, J.~Jaeckel, and D.~D. Lopes, {\it {Tracking axion-like
  particles at the LHC}},  \href{http://arxiv.org/abs/2302.12262}{{\tt
  arXiv:2302.12262}}.

\bibitem{Mitridate:2023tbj}
A.~Mitridate, M.~Papucci, C.~Wang, C.~Pe\~na, and S.~Xie, {\it {Energetic
  long-lived particles in the CMS muon chambers}},
  \href{http://arxiv.org/abs/2304.06109}{{\tt arXiv:2304.06109}}.

\bibitem{Dutta:2023abe}
B.~Dutta, D.~Kim, and H.~Kim, {\it {A Novel Beam-Dump Measurement with the LHC
  General-Purpose Detectors}},  \href{http://arxiv.org/abs/2305.16383}{{\tt
  arXiv:2305.16383}}.

\bibitem{Knapen:2016moh}
S.~Knapen, T.~Lin, H.~K. Lou, and T.~Melia, {\it {Searching for Axionlike
  Particles with Ultraperipheral Heavy-Ion Collisions}},  {\em Phys. Rev.
  Lett.} {\bf 118} (2017), no.~17 171801,
  [\href{http://arxiv.org/abs/1607.06083}{{\tt arXiv:1607.06083}}].

\bibitem{Knapen:2017ebd}
S.~Knapen, T.~Lin, H.~K. Lou, and T.~Melia, {\it {LHC limits on axion-like
  particles from heavy-ion collisions}},  {\em CERN Proc.} {\bf 1} (2018) 65,
  [\href{http://arxiv.org/abs/1709.07110}{{\tt arXiv:1709.07110}}].

\bibitem{CMS:2018erd}
{\bf CMS} Collaboration, A.~M. Sirunyan et~al., {\it {Evidence for
  light-by-light scattering and searches for axion-like particles in
  ultraperipheral PbPb collisions at $\sqrt{s_\mathrm{NN}} =$ 5.02 TeV}},  {\em
  Phys. Lett. B} {\bf 797} (2019) 134826,
  [\href{http://arxiv.org/abs/1810.04602}{{\tt arXiv:1810.04602}}].

\bibitem{ATLAS:2020hii}
{\bf ATLAS} Collaboration, G.~Aad et~al., {\it {Measurement of light-by-light
  scattering and search for axion-like particles with 2.2 nb$^{-1}$ of Pb+Pb
  data with the ATLAS detector}},  {\em JHEP} {\bf 11} (2021) 050,
  [\href{http://arxiv.org/abs/2008.05355}{{\tt arXiv:2008.05355}}].

\bibitem{SHiP:2015vad}
{\bf SHiP} Collaboration, M.~Anelli et~al., {\it {A facility to Search for
  Hidden Particles (SHiP) at the CERN SPS}},
  \href{http://arxiv.org/abs/1504.04956}{{\tt arXiv:1504.04956}}.

\bibitem{Berlin:2018pwi}
A.~Berlin, S.~Gori, P.~Schuster, and N.~Toro, {\it {Dark Sectors at the
  Fermilab SeaQuest Experiment}},  {\em Phys. Rev. D} {\bf 98} (2018), no.~3
  035011, [\href{http://arxiv.org/abs/1804.00661}{{\tt arXiv:1804.00661}}].

\bibitem{Bai:2021gbm}
Z.~Bai et~al., {\it {New physics searches with an optical dump at LUXE}},  {\em
  Phys. Rev. D} {\bf 106} (2022), no.~11 115034,
  [\href{http://arxiv.org/abs/2107.13554}{{\tt arXiv:2107.13554}}].

\bibitem{RebelloTeles:2023uig}
P.~Rebello~Teles, D.~d'Enterria, V.~P. Gon\c{c}alves, and D.~E. Martins, {\it
  {Searches for axion-like particles via $\gamma \gamma$ fusion at future
  $\mathrm{e}^+\mathrm{e}^-$ colliders}},
  \href{http://arxiv.org/abs/2310.17270}{{\tt arXiv:2310.17270}}.

\bibitem{EPICurl}
\url{https://www.jlab.org/conference/EPIC}.

\bibitem{Adkins:2022jfp}
J.~K. Adkins et~al., {\it Design of the ecce detector for the electron ion
  collider},  \href{http://arxiv.org/abs/2209.02580}{{\tt arXiv:2209.02580}}.

\bibitem{ATHENA:2022hxb}
J.~Adam and T.~A. collaboration, {\it Athena detector proposal — a totally
  hermetic electron nucleus apparatus proposed for ip6 at the electron-ion
  collider},  {\em Journal of Instrumentation} {\bf 17} (oct, 2022) P10019,
  [\href{http://arxiv.org/abs/2210.09048}{{\tt arXiv:2210.09048}}].

\bibitem{RHICurl}
\url{https://www.bnl.gov/rhic}.

\bibitem{STARurl}
\url{https://www.star.bnl.gov}.

\bibitem{Bock:2022lwp}
F.~Bock et~al., {\it {Design and Simulated Performance of Calorimetry Systems
  for the ECCE Detector at the Electron Ion Collider}},
  \href{http://arxiv.org/abs/2207.09437}{{\tt arXiv:2207.09437}}.

\bibitem{Kersevan:2004yh}
B.~P. Kersevan and E.~Richter-Was, {\it {Improved phase space treatment of
  massive multi-particle final states}},  {\em Eur. Phys. J. C} {\bf 39} (2005)
  439--450, [\href{http://arxiv.org/abs/hep-ph/0405248}{{\tt hep-ph/0405248}}].

\bibitem{Bern:2001dg}
Z.~Bern, A.~De~Freitas, L.~J. Dixon, A.~Ghinculov, and H.~L. Wong, {\it {QCD
  and QED corrections to light by light scattering}},  {\em JHEP} {\bf 11}
  (2001) 031, [\href{http://arxiv.org/abs/hep-ph/0109079}{{\tt
  hep-ph/0109079}}].

\bibitem{Klusek-Gawenda:2013rtu}
M.~Klusek-Gawenda and A.~Szczurek, {\it {$\pi^+ \pi^-$ and $\pi^0 \pi^0$ pair
  production in photon-photon and in ultraperipheral ultrarelativistic heavy
  ion collisions}},  {\em Phys. Rev. C} {\bf 87} (2013), no.~5 054908,
  [\href{http://arxiv.org/abs/1302.4204}{{\tt arXiv:1302.4204}}].

\bibitem{Diehl:2009yi}
M.~Diehl and P.~Kroll, {\it {Two-photon annihilation into octet meson pairs:
  Symmetry relations in the handbag approach}},  {\em Phys. Lett. B} {\bf 683}
  (2010) 165--171, [\href{http://arxiv.org/abs/0911.3317}{{\tt
  arXiv:0911.3317}}].

\bibitem{Ballam:1972eq}
J.~Ballam et~al., {\it {Vector Meson Production by Polarized Photons at
  2.8-GeV, 4.7-GeV, and 9.3-GeV}},  {\em Phys. Rev. D} {\bf 7} (1973) 3150.

\bibitem{Dolan:2017osp}
M.~J. Dolan, T.~Ferber, C.~Hearty, F.~Kahlhoefer, and K.~Schmidt-Hoberg, {\it
  {Revised constraints and Belle II sensitivity for visible and invisible
  axion-like particles}},  {\em JHEP} {\bf 12} (2017) 094,
  [\href{http://arxiv.org/abs/1709.00009}{{\tt arXiv:1709.00009}}]. [Erratum:
  JHEP 03, 190 (2021)].

\bibitem{dEnterria:2022sut}
D.~d'Enterria et~al., {\it {Opportunities for new physics searches with heavy
  ions at colliders}},  {\em J. Phys. G} {\bf 50} (2023), no.~5 050501,
  [\href{http://arxiv.org/abs/2203.05939}{{\tt arXiv:2203.05939}}].

\bibitem{Helm:1956zz}
R.~H. Helm, {\it {Inelastic and Elastic Scattering of 187-Mev Electrons from
  Selected Even-Even Nuclei}},  {\em Phys. Rev.} {\bf 104} (1956) 1466--1475.

\bibitem{Budnev:1975poe}
V.~M. Budnev, I.~F. Ginzburg, G.~V. Meledin, and V.~G. Serbo, {\it {The Two
  photon particle production mechanism. Physical problems. Applications.
  Equivalent photon approximation}},  {\em Phys. Rept.} {\bf 15} (1975)
  181--281.

\bibitem{Okun:1982xi}
L.~B. Okun, {\it {LIMITS OF ELECTRODYNAMICS: PARAPHOTONS?}},  {\em Sov. Phys.
  JETP} {\bf 56} (1982) 502.

\bibitem{Galison:1983pa}
P.~Galison and A.~Manohar, {\it {TWO Z's OR NOT TWO Z's?}},  {\em Phys. Lett.
  B} {\bf 136} (1984) 279--283.

\bibitem{Holdom:1985ag}
B.~Holdom, {\it {Two U(1)'s and Epsilon Charge Shifts}},  {\em Phys. Lett. B}
  {\bf 166} (1986) 196--198.

\bibitem{Pospelov:2007mp}
M.~Pospelov, A.~Ritz, and M.~B. Voloshin, {\it {Secluded WIMP Dark Matter}},
  {\em Phys. Lett. B} {\bf 662} (2008) 53--61,
  [\href{http://arxiv.org/abs/0711.4866}{{\tt arXiv:0711.4866}}].

\bibitem{Arkani-Hamed:2008hhe}
N.~Arkani-Hamed, D.~P. Finkbeiner, T.~R. Slatyer, and N.~Weiner, {\it {A Theory
  of Dark Matter}},  {\em Phys. Rev. D} {\bf 79} (2009) 015014,
  [\href{http://arxiv.org/abs/0810.0713}{{\tt arXiv:0810.0713}}].

\bibitem{Ilten:2016tkc}
P.~Ilten, Y.~Soreq, J.~Thaler, M.~Williams, and W.~Xue, {\it {Proposed
  Inclusive Dark Photon Search at LHCb}},  {\em Phys. Rev. Lett.} {\bf 116}
  (2016), no.~25 251803, [\href{http://arxiv.org/abs/1603.08926}{{\tt
  arXiv:1603.08926}}].

\bibitem{Abrahamyan:2011gv}
{\bf APEX} Collaboration, S.~Abrahamyan et~al., {\it {Search for a New Gauge
  Boson in Electron-Nucleus Fixed-Target Scattering by the APEX Experiment}},
  {\em Phys. Rev. Lett.} {\bf 107} (2011) 191804,
  [\href{http://arxiv.org/abs/1108.2750}{{\tt arXiv:1108.2750}}].

\bibitem{Lees:2014xha}
{\bf BaBar} Collaboration, J.~P. Lees et~al., {\it {Search for a Dark Photon in
  $e^+e^-$ Collisions at BaBar}},  {\em Phys. Rev. Lett.} {\bf 113} (2014),
  no.~20 201801, [\href{http://arxiv.org/abs/1406.2980}{{\tt
  arXiv:1406.2980}}].

\bibitem{Adrian:2018scb}
P.~H. Adrian et~al., {\it {Search for a Dark Photon in Electro-Produced
  $e^{+}e^{-}$ Pairs with the Heavy Photon Search Experiment at JLab}},
  \href{http://arxiv.org/abs/1807.11530}{{\tt arXiv:1807.11530}}.

\bibitem{Anastasi:2015qla}
A.~Anastasi et~al., {\it {Limit on the production of a low-mass vector boson in
  $\mathrm{e}^{+}\mathrm{e}^{-} \to \mathrm{U}\gamma$, $\mathrm{U} \to
  \mathrm{e}^{+}\mathrm{e}^{-}$ with the KLOE experiment}},  {\em Phys. Lett.}
  {\bf B750} (2015) 633--637, [\href{http://arxiv.org/abs/1509.00740}{{\tt
  arXiv:1509.00740}}].

\bibitem{Anastasi:2018azp}
{\bf KLOE-2} Collaboration, A.~Anastasi et~al., {\it {Combined limit on the
  production of a light gauge boson decaying into $\mu^+\mu^-$ and
  $\pi^+\pi^-$}},  {\em Submitted to: Phys. Lett. B} (2018)
  [\href{http://arxiv.org/abs/1807.02691}{{\tt arXiv:1807.02691}}].

\bibitem{Aaij:2019bvg}
{\bf LHCb} Collaboration, R.~Aaij et~al., {\it {Search for
  $A'\!\to\!\mu^+\mu^-$ decays}},  \href{http://arxiv.org/abs/1910.06926}{{\tt
  arXiv:1910.06926}}.

\bibitem{Batley:2015lha}
{\bf NA48/2} Collaboration, J.~R. Batley et~al., {\it {Search for the dark
  photon in $\pi^0$ decays}},  {\em Phys. Lett.} {\bf B746} (2015) 178--185,
  [\href{http://arxiv.org/abs/1504.00607}{{\tt arXiv:1504.00607}}].

\bibitem{Tsai:2019mtm}
Y.-D. Tsai, P.~deNiverville, and M.~X. Liu, {\it {The High-Energy Frontier of
  the Intensity Frontier: Closing the Dark Photon, Inelastic Dark Matter, and
  Muon g-2 Windows}},  \href{http://arxiv.org/abs/1908.07525}{{\tt
  arXiv:1908.07525}}.

\bibitem{Andreas:2012mt}
S.~Andreas, C.~Niebuhr, and A.~Ringwald, {\it {New Limits on Hidden Photons
  from Past Electron Beam Dumps}},  {\em Phys. Rev. D} {\bf 86} (2012) 095019,
  [\href{http://arxiv.org/abs/1209.6083}{{\tt arXiv:1209.6083}}].

\bibitem{Konaka:1986cb}
A.~Konaka et~al., {\it {Search for Neutral Particles in Electron Beam Dump
  Experiment}},  {\em Phys. Rev. Lett.} {\bf 57} (1986) 659.

\bibitem{Banerjee:2019hmi}
D.~Banerjee et~al., {\it {Improved limits on a hypothetical X(16.7) boson and a
  dark photon decaying into $e^+e^-$ pairs}},
  \href{http://arxiv.org/abs/1912.11389}{{\tt arXiv:1912.11389}}.

\bibitem{Astier:2001ck}
{\bf NOMAD} Collaboration, P.~Astier et~al., {\it {Search for heavy neutrinos
  mixing with tau neutrinos}},  {\em Phys. Lett.} {\bf B506} (2001) 27--38,
  [\href{http://arxiv.org/abs/hep-ex/0101041}{{\tt hep-ex/0101041}}].

\bibitem{Davier:1989wz}
M.~Davier and H.~Nguyen~Ngoc, {\it {An Unambiguous Search for a Light Higgs
  Boson}},  {\em Phys. Lett.} {\bf B229} (1989) 150--155.

\bibitem{Bernardi:1985ny}
G.~Bernardi et~al., {\it {Search for Neutrino Decay}},  {\em Phys. Lett.} {\bf
  166B} (1986) 479--483.

\bibitem{Bjorken:2009mm}
J.~D. Bjorken, R.~Essig, P.~Schuster, and N.~Toro, {\it {New Fixed-Target
  Experiments to Search for Dark Gauge Forces}},  {\em Phys. Rev. D} {\bf 80}
  (2009) 075018, [\href{http://arxiv.org/abs/0906.0580}{{\tt
  arXiv:0906.0580}}].

\bibitem{Liu:2017htz}
Y.-S. Liu and G.~A. Miller, {\it {Validity of the Weizs\"acker-Williams
  approximation and the analysis of beam dump experiments: Production of an
  axion, a dark photon, or a new axial-vector boson}},  {\em Phys. Rev. D} {\bf
  96} (2017), no.~1 016004, [\href{http://arxiv.org/abs/1705.01633}{{\tt
  arXiv:1705.01633}}].

\end{thebibliography}\endgroup

\end{document}